\documentclass[11pt,a4paper,twoside,openright,closeany,textany]{book}
\usepackage{makeidx}
\usepackage{graphicx}
\usepackage{paralist}
\usepackage{caption} 
\usepackage{amsmath}
\usepackage{amssymb}
\usepackage{mathtools}
\usepackage{fancyhdr}
\usepackage{multicol}
\usepackage{multirow}

\usepackage{mdframed}
\usepackage[usenames,dvipsnames,svgnames,table]{xcolor}
\usepackage{array}
\usepackage[%
  colorlinks=true,%
  urlcolor=blue,%
  citecolor=blue%
]{hyperref}
\usepackage{nameref} 
\usepackage{ifthen} 
\usepackage[font={small,sl}]{caption}
\usepackage[authoryear]{natbib}
\renewcommand\bibname{References}
\newcommand{\mychapbib}{
  \addcontentsline{toc}{section}{\bibname}
  \bibliographystyle{natbib}
  \bibliography{strucbioinf}
}
\usepackage[colorinlistoftodos]{todonotes}
\usepackage{letltxmacro}

\usepackage{tocstyle} 
\usetocstyle{standard}
\settocfeature{raggedhook}{\raggedright}
\newcommand{\bibfont}{\small\raggedright}
\def\cite{\citep}

\LetLtxMacro{\oldTodo}{\todo}
\renewcommand{\todo}[2][]{\oldTodo[#1]{TODO: #2}}
\newcommand\intodo[1]{\todo[inline]{#1}}
\usepackage[normalem]{ulem}

\newcommand\inwish[1]{\oldTodo[inline,color=SkyBlue]{WISH: #1}}

\newboolean{onechapter}

\newcommand{\AF}{K.\@~Anton~Feenstra}
\newcommand{\SA}{Sanne~Abeln}

\newcommand{\HM}{Halima~Mouhib}

\newcommand{\BS}{Bas~Stringer}

\newcommand{\JG}[1][~]{\mbox{Jose}#1\mbox{Gavald\'a-Garc\'ia}}

\newcommand{\AR}[1][~]{\mbox{Arri\"en}#1\mbox{Symon}#1\mbox{Rauh}}

\newcommand{\KW}[1][~]{\mbox{Katharina}#1\mbox{Waury}}
\newcommand{\HI}[1][~]{\mbox{Hugo}#1\mbox{van}#1\mbox{Ingen}}

\newcommand{\orcid}[1]{\href{https://orcid.org/#1}{\raisebox{-0.7ex}{\protect\includegraphics[height=3ex]{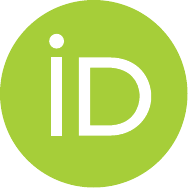}}}}
\definecolor{idgreen}{RGB}{166 206 57}
\newcommand{\mailid}[1]{\href{mailto:#1}{\raisebox{-0.3ex}{\color{idgreen}\textsf{\textbf{\Large \protect@}}}}}

\newcommand{\AFid}{\orcid{0000-0001-6755-9667}}
\newcommand{\SAid}{\orcid{0000-0002-2779-7174}}
\newcommand{\HMid}{\orcid{0000-0001-5031-3468}}
\newcommand{\JGid}{\orcid{0000-0001-6431-3442}}

\newcommand{\ARid}{\orcid{0000-0001-9707-3836}}

\newcommand{\KWid}{\orcid{0000-0002-8570-7640}}

\newcommand{\BSid}{\orcid{0000-0001-7792-9385}}

\newcommand{\HIid}{\orcid{0000-0002-0808-3811}}

\newcommand{\ACtxt}{Wrote the text}
\newcommand{\ACfig}{Created figures}
\newcommand{\ACref}{Review of current literature}
\newcommand{\ACeds}{Editorial responsibility}
\newcommand{\ACproof}{Critical proofreading}
\newcommand{\ACfb}{Non-expert feedback}

\newcommand{\Angs}[1][~]{\text{\normalfont\AA}}

\renewcommand{\and}{\quad}

\newcommand{\pdbref}[1]{\href{http://www.rcsb.org/pdb/explore.do?structureId=#1}{PDB:#1}}
\newcommand{\arxiv}[2][UNDEFINED]{\href{https://arxiv.org/abs/#2}{\ifthenelse{\equal{#1}{UNDEFINED}}{arxiv.org/abs/#2}{#1}}}

\newcommand{\figref}[2][]{\hyperref[fig:#2]{Figure\@~\ref*{fig:#2}#1}}
\newcommand{\tabref}[1]{\hyperref[tab:#1]{Table \ref*{tab:#1}}}
\renewcommand{\eqref}[1]{\hyperref[eq:#1]{Equation\@~\ref*{eq:#1}}}
\newcommand{\panelref}[2][]{%
    \ifthenelse{\boolean{onechapter}}{%
        \hyperref[panel:#2]{Panel\@~``\nameref{panel:#2}#1''}%
    }{%
        \hyperref[panel:#2]{Panel\@~\ref*{panel:#2}#1}%
    }%
}
\newcommand{\secref}[1]{\hyperref[sec:#1]{Section\@~\ref*{sec:#1}}}
\newcommand{\chref}[2][n]{%
    \ifthenelse{\boolean{onechapter}}{%
        \ifthenelse{\equal{#2}{ChPref}     }{\arxiv[Chapter ``\nameref*{ch:#2}'']{1801.09442}}{}%
        \ifthenelse{\equal{#2}{ChIntroPS}  }{\arxiv[Chapter ``\nameref*{ch:#2}'']{1801.09442}}{}%
        \ifthenelse{\equal{#2}{ChDetVal}   }{\arxiv[Chapter ``\nameref*{ch:#2}'']{1801.09442}}{}%
        \ifthenelse{\equal{#2}{ChStrucAli} }{\arxiv[Chapter ``\nameref*{ch:#2}'']{1801.09442}}{}%
        \ifthenelse{\equal{#2}{ChDBClass}  }{\arxiv[Chapter ``\nameref*{ch:#2}'']{1801.09442}}{}%
        \ifthenelse{\equal{#2}{ChFunc}     }{\arxiv[Chapter ``\nameref*{ch:#2}'']{1801.09442}}{}%
        \ifthenelse{\equal{#2}{ChIntroPred}}{\arxiv[Chapter ``\nameref*{ch:#2}'']{1712.00407}}{}%
        \ifthenelse{\equal{#2}{ChHomMod}   }{\arxiv[Chapter ``\nameref*{ch:#2}'']{1712.00425}}{}%
        \ifthenelse{\equal{#2}{ChSSPred}   }{\arxiv[Chapter ``\nameref*{ch:#2}'']{1801.09442}}{}%
        \ifthenelse{\equal{#2}{ChFuncPred} }{\arxiv[Chapter ``\nameref*{ch:#2}'']{1801.09442}}{}%
        \ifthenelse{\equal{#2}{ChIntroDyn} }{\arxiv[Chapter ``\nameref*{ch:#2}'']{1801.09442}}{}%
        \ifthenelse{\equal{#2}{ChThermo}   }{\arxiv[Chapter ``\nameref*{ch:#2}'']{1801.09442}}{}%
        \ifthenelse{\equal{#2}{ChMD}       }{\arxiv[Chapter ``\nameref*{ch:#2}'']{1801.09442}}{}%
        \ifthenelse{\equal{#2}{ChMC}       }{\arxiv[Chapter ``\nameref*{ch:#2}'']{1801.09442}}{}%
    }{
    \hyperref[ch:#2]{%
        \ifthenelse{\equal{#1}{n} }{Chapter \ref*{ch:#2}}{}% just number
        \ifthenelse{\equal{#1}{nn}}{Chapter \ref*{ch:#2} ``\nameref{ch:#2}''}{}% name & number
        \ifthenelse{\equal{#1}{N} }{``\nameref{ch:#2}''}{}% just name
      }%
  }%
}
\newcommand{\chrefname}[1]{\hyperref[ch:#1]{Chapter \ref*{ch:#1} ``\nameref{ch:#1}''}}
\newcommand{\partref}[1]{\hyperref[#1]{Part \ref*{#1}}}
\newcommand{\appref}[1]{\hyperref[app:#1]{Appendix \ref*{app:#1}}}

\newcommand{\figsource}[1]{\protect\footnote{Figure source location: \url{#1}}}

\renewcommand{\arraystretch}{1.3}

\makeatletter \@beginparpenalty=5000 \makeatother

\newenvironment{panel}[1][]{
  \begin{figure}[htb]
    \begin{mdframed}[%
        outerlinewidth=0,%
        linecolor=CornflowerBlue!30,%
        backgroundcolor=CornflowerBlue!30,%
        innerleftmargin=14,%
        innerrightmargin=14,%
      ]
      \ifthenelse{\equal{#1}{}}{}{
        \stepcounter{panel}
		\subsection*{#1} 
      }
}{%
    \end{mdframed}
  \end{figure}
}

\newenvironment{bgreading}[1][]{
  \begin{mdframed}[%
      outerlinewidth=0,%
      linecolor=CornflowerBlue!30,%
      backgroundcolor=CornflowerBlue!30,%
      innerleftmargin=14,%
      innerrightmargin=14,%
    ]
	\ifthenelse{\equal{#1}{}}{}{
        \stepcounter{panel}
    	\subsection*{#1} 
    }
}{%
  \end{mdframed}
}

\usepackage{listings}
\definecolor{backcolour}{rgb}{0.95,0.95,0.92}
\definecolor{codegreen}{rgb}{0,0.6,0}
\definecolor{codegray}{rgb}{0.5,0.5,0.5}
\definecolor{codered}{rgb}{0.8,0,0.0}
\definecolor{codeblue}{rgb}{0.0,0,0.8}
\lstdefinestyle{codeStyle}{
    backgroundcolor=\color{backcolour},   
    commentstyle=\color{codegreen},
    keywordstyle=\color{codeblue},
    numberstyle=\tiny\color{codegray},
    stringstyle=\color{codegray},
    numbers=left,                    
    tabsize=2
} 
\makeindex

\begin{document}

\setboolean{onechapter}{true}

\pagestyle{fancy}
\lhead[\small\thepage]{\small\sf\nouppercase\rightmark}
\rhead[\small\sf\nouppercase\leftmark]{\small\thepage}
\newcommand{\innerfoot}{\footnotesize{\sf{\copyright} Feenstra \& Abeln}, 2014-2021}
\newcommand{\outerfoot}{\footnotesize \sf Intro Struc Bioinf}
\lfoot[\outerfoot]{\innerfoot}
\cfoot{}
\rfoot[\innerfoot]{\outerfoot}
\renewcommand{\footrulewidth}{\headrulewidth}

\mainmatter

\setcounter{chapter}{1}
\chapterauthor{\HM~\HMid \and \BS~\BSid \and \HI~\HIid \and \JG~\JGid \and \KW~\KWid \and \SA~\SAid \and \AF~\AFid}
\chapterfigure{\includegraphics[width=0.5\linewidth]{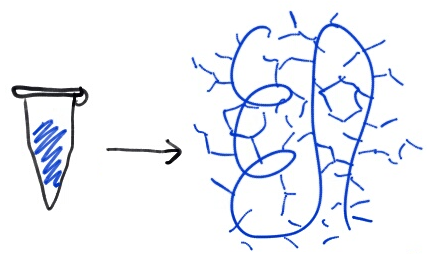}}
\chapter{Structure determination}
\label{ch:ChDetVal}

\ifthenelse{\boolean{onechapter}}{\tableofcontents\newpage}{}


\section{Introduction}
\begin{figure}[b]
\includegraphics[width=\linewidth]{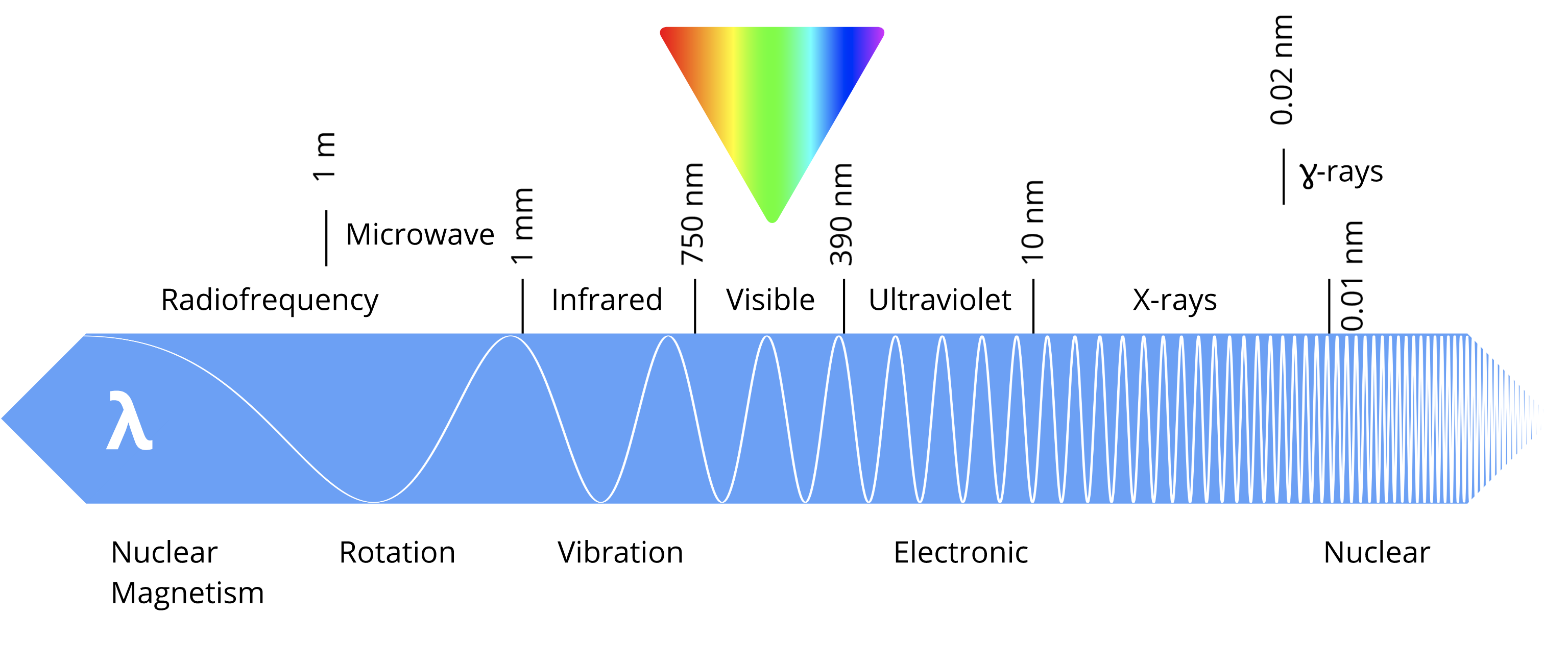}
\caption{Electromagnetic waves spectrum and their applications. In principle all (or most) wavelengths of the electromagnetic spectrum can be used to obtain information from molecules. Different kinds of electromagnetic waves (top) are used to obtain diverse information on molecular systems (bottom). 
}
\figsource{https://drive.google.com/open?id=1enROXCNtLG5zrzvOTlqVtWQlR221_-S-}
\label{fig:ChStrucDet-em-spectrum}
\end{figure}

The main emphasis of this work is to provide a background on experimental techniques for protein structure determination. The focus is set on X-ray crystallography and Nuclear Magnetic Resonance spectroscopy (NMR), which are by far the main methods used to determine the structure of soluble proteins. We will also introduce cryogenic Electron Microscopy (cryo-EM) and electron diffraction which are more suited to analyze membrane proteins and larger protein complexes. At the end, more qualitative techniques are summarized that are used to obtain insight on the overall structure and dynamics of proteins. Note that this introduction to protein structure determination  aims at familiarizing the reader to different experimental techniques, their benefits and bottlenecks, but that a thorough mathematical and technical description of the concept is beyond the scope of this work. For the interested reader, \secref{ChDetVal:reading} provides selected works that go deeper into the details. 

Generally speaking, structure determination is an immediate result of understanding the interaction between light (radiation) and matter. \figref{ChStrucDet-em-spectrum} shows an overview of the electromagnetic radiation (e.g.\@ light) and the corresponding wavelengths used by the different experimental techniques. A corresponding overview of the methods including the used wavelengths, their reachable resolutions, benefits, and limitations is given in \tabref{ChDetVal-overview}. Depending on the sample of interest, different techniques are applied to resolve the structure. 

\begin{table}[h]
\caption{Overview of different Methods used for protein structure determination.}
\label{tab:ChDetVal-overview}
\centerline{
\begin{tabular}{ccccc}
\hline
\textbf{Method}&\textbf{Resolution} 
                     &\textbf{Wavelengths}&\textbf{Strengths}
                                                        &\textbf{Limitations}\\
\hline
\textbf{X-ray}  & $1-2$\AA    & nm-pm         & atomic resolution & purity \& crystal     \\[-1ex]
                &           &               &                   & static     \\
\textbf{NMR}    & $1-2$\AA    & m       & is in solution    & only small proteins   \\[-1ex]
                &           &               & dynamics          &            \\[-1ex]
                &           &               & direct interactions          &            \\
\textbf{EM}     & mm-$\mu$m & mm-$\mu$m$^*$ & direct imaging    & very low resolution   \\[-1ex]
                &           &               & unpurified        & static     \\
\textbf{Cryo-EM}&$\sim2-10$\AA$\mu$m-nm$^*$& measure phases  & not atomic resolution \\[-1ex]
                &           &               & large complexes   & static     \\
\textbf{IR}     & --        & mm-$\mu$m     & global structure  & no atomic assignments \\[-1ex]
                &           &               & dynamics          &            \\
\textbf{CD}     & --        & $10-100$nm      & global  structure & no atomic assignments \\[-1ex]
                &           &               & dynamics          &            \\
\hline
\end{tabular}
}
X-ray: protein X-ray crystallography; NMR: Nuclear Magnetic Resonance spectroscopy; EM: Electron microscopy; IR: Infrared spectroscopy; CD: Circular Dichroism; 
$^*$ electron radiation. \\
Note: Atomic resolution lies in the range of 1-2 {\Angs}: the atomic (Van der Waals) radius of carbon is about 1.5 {\Angs}, that of nitrogen and oxygen 1.1 {\Angs}. Bond lengths between carbon, nitrogen or oxygen are in that same range.
\end{table}

X-ray Crystallography and NMR have traditionally been the two main methods for protein structure determination, as they can obtain the highest `atomic' resolutions -- in the range of 1-2 {\Angs}. X-ray crystallography uses short wavelength X-ray radiation; for X-ray, the wavelengths used limit the resolution of the diffraction data obtained. Shorter wavelengths give higher resolution information. 

NMR uses the long wavelength radio waves as these frequencies correspond to the energy levels of spin-state transitions in the nuclei of atoms, which are sensitive to the local (atomic) environment. This local environment yields information about the relative positions of atoms, from which the overall protein structure is constructed. 

Over the past decades, cryo-EM has been steadily pushing down on the resolution limit, going down from 4 {\Angs} in 2008 \cite{Yu2008,Zhang2008}, over ``near-atomic'' resolutions of 3-3.5 {\Angs} \cite{Li2013a,Earl2017} to atomic resolution close to 1 {\Angs} \cite{Herzik2020}. This technique is particularly interesting for large assemblies such as viruses \cite{Jiang2017,Ward2017} and flagella \cite{Egelman2017}.

All successfully resolved and published structures, usually obtained from X-ray, NMR or cryo-EM, are accessible via the Protein Data Bank (PDB; \url{www.pdb.org}, see \chref{ChDBClass} for more detail). 
Besides X-ray, NMR and cryo-EM, which yield direct information on atomic coordinates, other spectroscopic techniques (visible, UV, IR) are sensitive to electronic and molecular vibrations. These measurements can be used to probe various properties of molecular systems and obtain more global and qualitative information (e.g.\@ the amount of secondary structure elements). We will introduce these methods in some more detail throughout the chapter to provide a general overview of available techniques in structure determination as well as sufficient references to dig in deeper yourselves.

\section{X-ray crystallography}

\begin{figure}[h]
\centerline{
\includegraphics[width=1.1\linewidth]{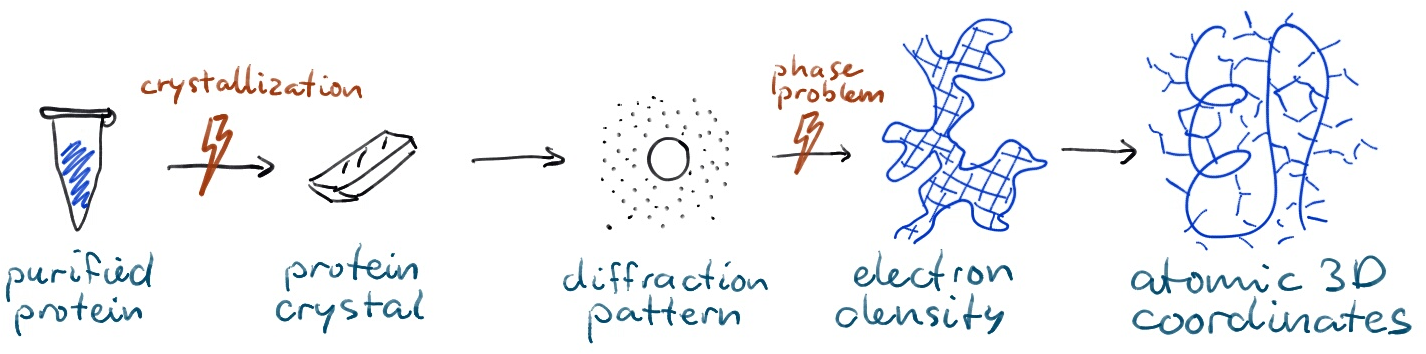}}
\caption{Simplified work-flow of protein structure determination through X-ray crystallography. Crystallization and the phase problem are the main bottlenecks.}
\label{fig:ChStrucDet-xray}
\figsource{https://drive.google.com/open?id=1D7BcdbLzZ3xj3IWvZzQ7PyWqvI0_lM-o}
\end{figure}

A simplified work-flow used in X-ray crystallography, which to this date is still the method that provided most of the available protein structures, is shown in \figref{ChStrucDet-xray}. First, the proteins need to form a regular crystal structure, such that their orientation is regular and very densely packed against each other. The fixed orientation of the crystals allows the X- ray diffraction, in the next step, to be recorded as a regular pattern. In a third step the electron density may be acquired from the diffraction pattern, but first the phase problem needs to be solved. As a result, a 3D structure may be fitted to the density. The different steps and relevant concepts are explained in the following sections. Note that obtaining good crystals of the protein, and the derivation of the electron density from the diffraction data are the more challenging parts (highlighted with red lightning bolts in \figref{ChStrucDet-xray}).

\subsection{Crystallization}
\begin{figure}
\centerline{
\textsf{A}\includegraphics[height=0.45\linewidth]{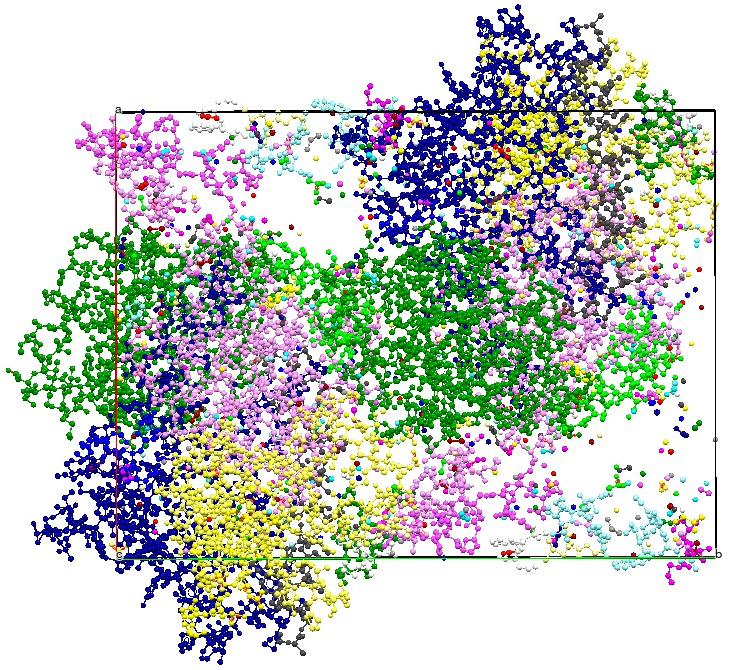}
\textsf{B}\includegraphics[height=0.45\linewidth]{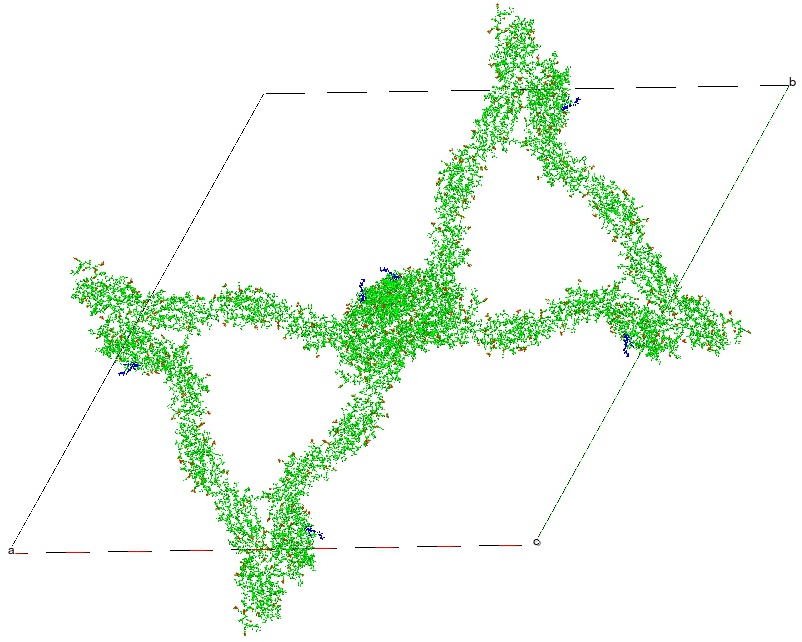}
}
\caption{A) A typical protein crystal of human deoxyhemoglobin (\pdbref{4hhb}) with a relatively low fraction of water. (B) The crystal packing of Myelin-associated glycoprotein (\pdbref{5lf5}), which contains an exceptional amount of water. The elementary cell (repetitive units in the crystal) are outlined with rectangles. }
\label{fig:ChStrucDet-crystal}
\end{figure}

When you try to imagine a crystal, you may first think about something like salt or sugar grains, or some rock or gem. While protein crystals have a similar appearance, inside, they contain a surprisingly large amount of water: between 20\% and up to as much as 80\% by volume. \figref{ChStrucDet-crystal} shows examples of a particularly dense (little water) and an open (much water) crystal elementary cell.
You should realize that, for example, a regular packing of spheres (think of a box of marbles) also contains about 25\% empty space (see also \citet{Atkins2002} on crystal packing); however 80\% water is more like a gel than a crystal. This is not so different compared to the typical cytosol (inside a living cell), which is about 70\% water \cite{Luby-Phelps1999} with most of the rest (20-30\%) taken up by proteins \cite{Ellis2001}. As one might expect, anything that makes a protein flexible can interfere with the crystallization process. This problem is two-fold: not all proteins will be in the same conformation, making it difficult to obtain the regular packing required in the crystal. (Although in a few cases, a single crystal may contain different conformations of the same protein -- this happens already naturally in some virus capsids.) The second part of the problem is related to the loss of entropy upon crystallization, since the protein molecules become more ordered as they occupy their fixed positions on the crystal lattice (we will return to entropic effects in \chref{ChThermo}). 

\begin{bgreading}[Challenging structures]
\label{panel:ChDetVal:challenging}
\begin{compactitem}
\item[\textbf{Membrane proteins}] such as G protein coupled receptors (GPCRs) in particular pose a challenge. These protein represent about 60\% of the drug targets and are thus extremely important for drug design. They need to be embedded within the membrane to be stable, but the membrane consists of many small (lipid) molecules and is very flexible, which makes it almost impossible to fit into a crystal. Even though a whole array of tricks has been invented (like using simple detergents in stead of lipids; or even inducing two-dimensional crystallization inside a membrane), this is still the largest bottleneck in protein crystallography. Until 2005 the only available crystal structures of GPCRs were of rhodopsin. Also, until today, it is still not possible to crystallize olfactory GPCRs which are responsible for detecting odorants in the nose.

\item[\textbf{Glycoproteins}] are another example that pose a double challenge for crystallographers: first, the glycan (sugar) groups are very flexible, which like flexible linkers, loops or termini, interferes with crystallization. But the (even) greater problem is heterogeneity. The glycan groups are added after translation, i.e.\@ so-called post-translational modifications. This is done by enzymes which have specific affinities for attaching certain glycans in given places on the protein. But the placing, number and types attached may vary from protein molecule to protein molecule. This means the crystal must now accommodate protein molecules which have slightly different shapes. You can imagine that this will not work very well. Moreover, without the glycans attached, many of these proteins adopt different conformations, or even remain largely disordered. And, finally, the enzyme machinery for attaching the glycans can vary between species, and only eucaryotes have them. This makes production of these proteins in the right form experimentally challenging as well. 
\end{compactitem}
New developments in cryo-electron microscopy (cryo-EM) will allow to obtain more information on these kind of challenging systems. Since the determination of the first EM structure of an activated GPCR at approximately 4 {\Angs} resolution in 2017 \cite{Zhang2017c}, cryo-EM has very recently moved towards atomic resolution close to 1 {\Angs} and is even expected to reach resolutions below 1 {\Angs} in the years to come \cite{Herzik2020}. We will come back to cryo-EM in \secref{ChDetVal:cryoEM}. 

\end{bgreading}

\subsection{Diffraction}

In an X-ray diffraction experiment, a well-defined and narrow X-ray beam is directed at the crystal. Usually some tungsten based X-ray generator is used as the radiation source to provide light of a given wavelength between 0.1-10 nm (see also \figref{ChStrucDet-em-spectrum}).
During the data acquisition, the crystal is often cooled by so-called `cryogen': a stream of liquid nitrogen or helium . This has two reasons. First, the X-ray radiation hitting the crystal will heat it up and eventually cause damage. Second, to obtain highest resolutions, atomic motions have to be reduced.

The radiation that is diffracted by the crystal is then captured and recorded by a detector (all radiation that goes straight through the crystal on the other hand is stopped by a small slab of metal called the ``beam stop''). The diffracted radiation makes up a diffraction pattern which contains the information that we need to derive the coordinates of the atoms in our protein.

\begin{panel}[Constructive and destructive interference]
\label{panel:ChStrucDet-waves}
\includegraphics[width=\linewidth]{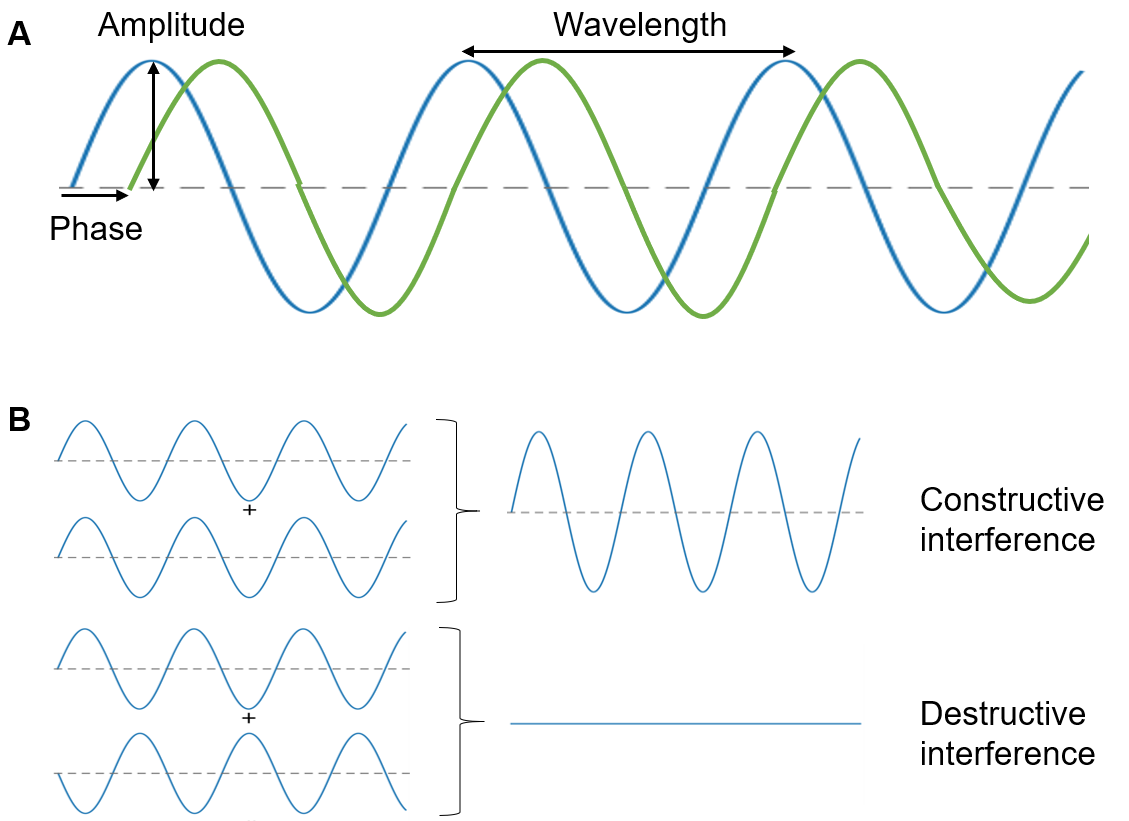}
X-ray crystallography works because the waves are phase-shifted while passing through the crystal, and the superposition of outgoing waves leads to the diffraction pattern. A: Definition of phase shift, amplitude, and wavelength. B: Interference of two waves in phase (upper trace) and 180 degree out of phase. While the former leads to constructive interference and a wave of larger amplitude, the latter results in destructive interference and the destruction of the two waves.
\end{panel}
It is the easiest to envision the process for a single protein atom at a time: the incoming radiation will strike the atoms in each of the protein molecules in the crystal. Because the atoms are in different positions, the radiation for each atom will travel its own specific distance (``path length'') from the source to the detector.
Each radiation wave is defined by two properties: its amplitude and its phase (see \panelref{ChStrucDet-waves}~A).
For different directions, the waves will be effectively randomized, and most energy gets averaged away or even canceled out (destructive interference). However, in some specific directions the waves are in phase, so the waves will be amplified, and the intensities add up (constructive interference) to produce a spot on the detector. Both types of interference are shown in \panelref{ChStrucDet-waves}~B. Each different atom in the protein produces multiple such spots, but many different atoms may also contribute to the same spot. The relation to describe the path length differences as a multiple of the wavelength, n$\lambda$, is known as Bragg's law in crystallography:

\begin{equation}
\ 2d\cdot sin \theta = n\lambda
\end{equation}

\begin{figure}[h]
\includegraphics[width=\linewidth]{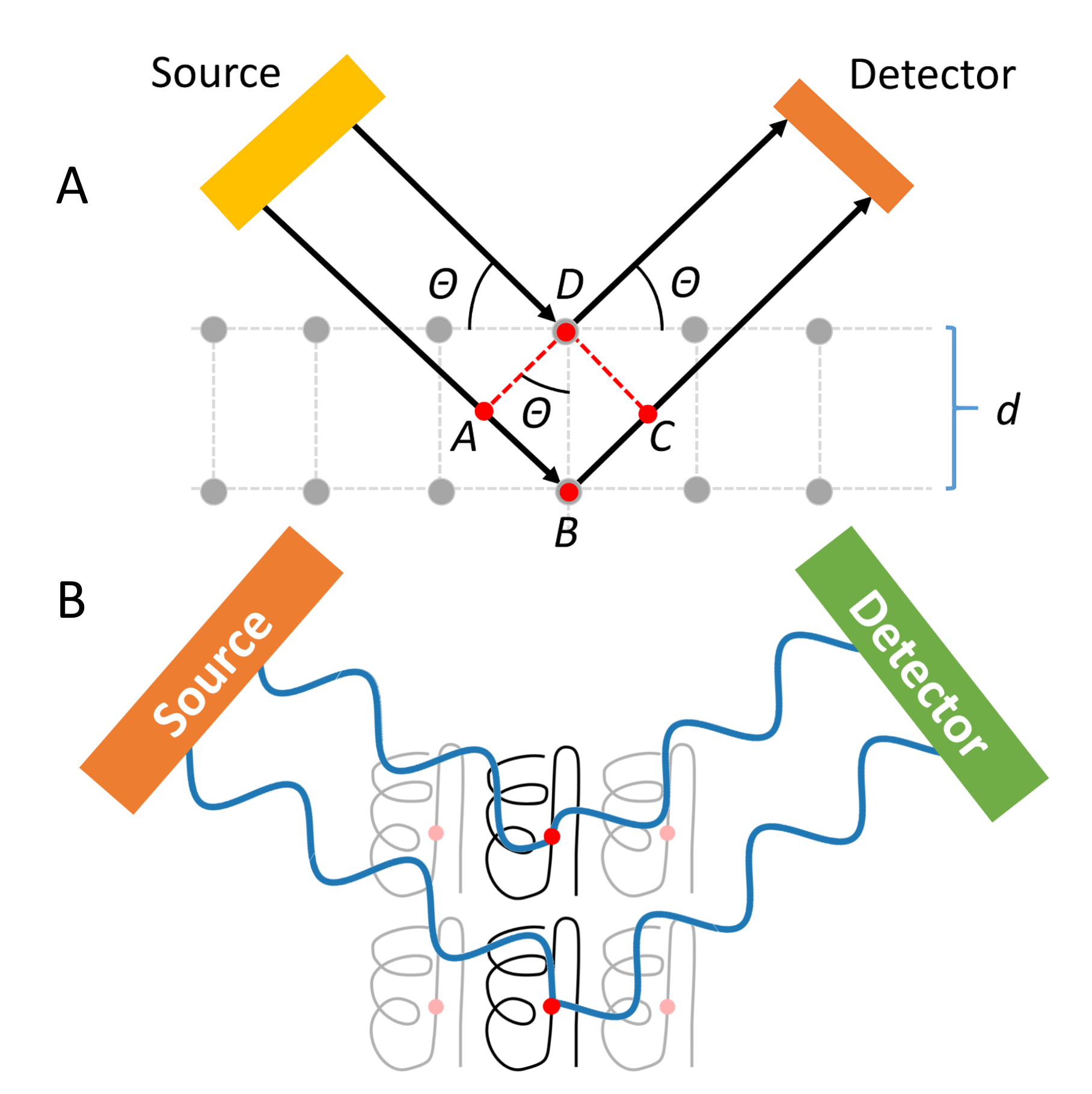}
\caption{Bragg's law. A: Simplified scheme to define the distance d between two planes of the crystal lattice, the glancing angle theta, wavelength lambda, and the diffraction order n using 4 atoms (A, B, C, D) in a crystal lattice. B: Put into context of atoms inside the protein molecules of the crystal. Incident radiation is drawn to come from the left. For simplification, we show two atoms in two different lattice layers, that scatter the radiation in a specific angle onto the detector.}
\label{fig:ChStrucDet-bragg}
\end{figure}

Hereby, $d$ represents the distance between two planes in the crystal, $\theta$ (theta) is the glancing angle, lambda the wavelength, and n the diffraction order.  \figref{ChStrucDet-bragg} shows Bragg's law in context of protein molecules inside a crystal. Considering basic trigonometric rules, you can easily derive the relations between the angles and lengths to determine the distance $d$ between the planes of the crystal. Note that constructive interference will only be achieved if Bragg's law is fulfilled.

\begin{bgreading}[Diffraction Pattern]
A typical diffraction pattern is shown below. The amount of detail, or resolution, of the data increases with distance from the centre. Thus, the crystallographer can immediately say what the maximum resolution could be, by looking at the furthest observed `reflections' -- the black dots scattered in patterns across the image (highlighted in red below). It is important to note that the diffraction pattern is the (only) primary data that an X-ray experiment produces. All the rest (densities, atomic coordinates, B-factors), are modelled onto the primary data in one way or another.

\centerline{\includegraphics[width=0.6\linewidth]{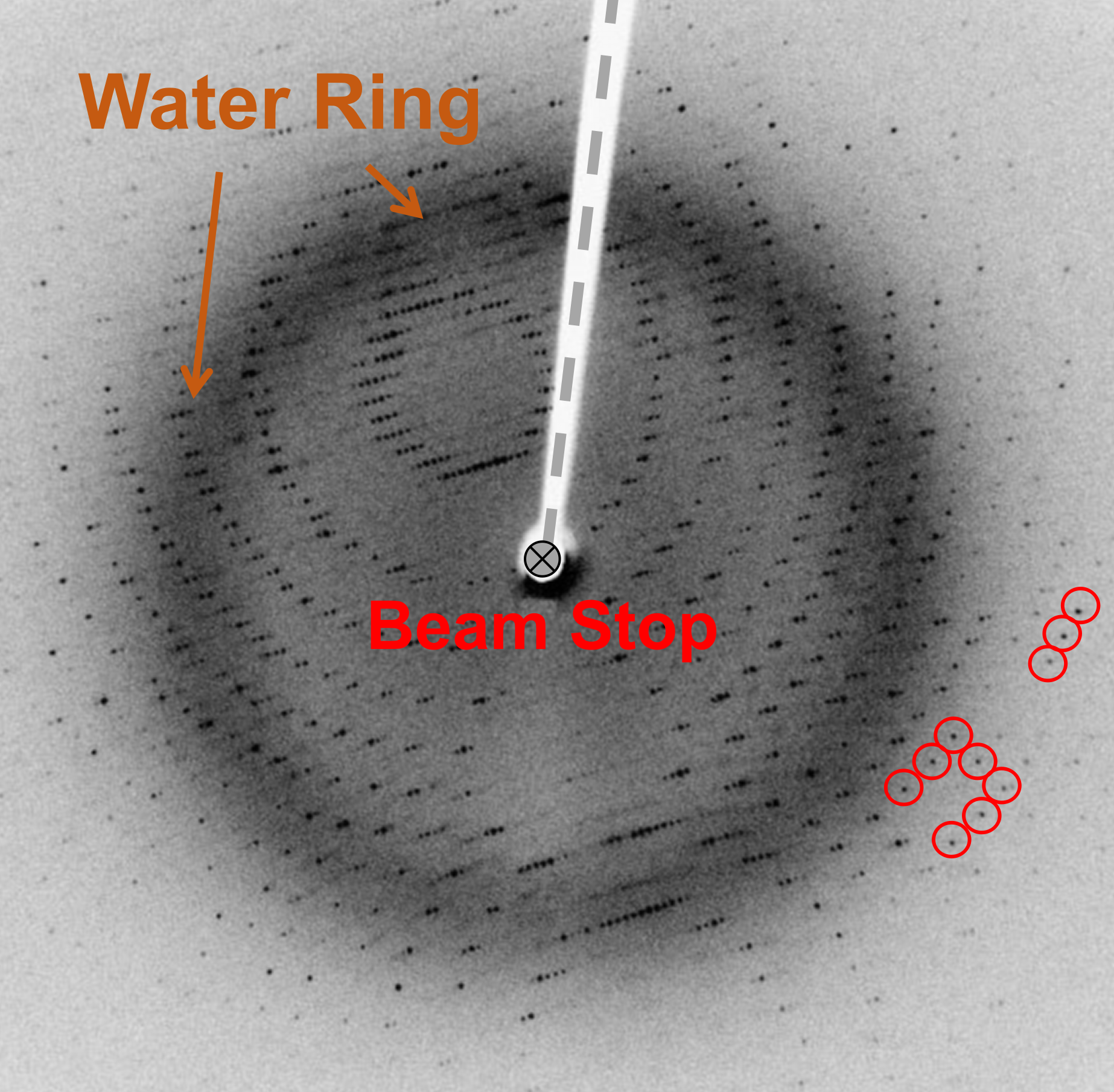}}
\noindent\textsl{X-ray diffraction pattern of crystallized 3Clpro, a SARS protease (2.1 {\AA} resolution) \footnote{Source: Jeff Dahl \url{https://commons.wikimedia.org/wiki/File:X-ray_diffraction_pattern_3clpro.jpg}}.}

The diffraction pattern has several striking features that do not carry information about the protein structure. In the middle is a blank area, caused by the beam stop preventing any (`direct beam') radiation from reaching the detector. Unfortunately, protein crystals only diffract a (small) fraction of the incoming radiation, so not using a beam stop would completely overwhelm the detector (akin to pointing a camera at the sun). There will be a ring caused by diffraction on the randomly oriented water in the crystal, known as the `water ring'. The intensity of this ring will depend on the fraction of water present. Also the loop or, in this case, the rod that holds the crystal will scatter some of the radiation (not diffraction, just bouncing off the surface).

Finally, we see many small dots known as reflections. You can see they lie in a pattern, which will vary depending on the type (symmetry) of the arrangements (packing) of the protein molecules in the crystal. In this lattice pattern, each point has a set of three indices, relating to the angles of the diffraction. The actual data used are the intensities of each (observed) spot at each possible lattice (index) position.
\end{bgreading}

Mathematically, since the atoms are regularly arranged in the crystal, the observed pattern now corresponds to a (three dimensional) Fourier transform 
of the positions of the atoms. So, in principle we would just need to do a reverse Fourier transform to obtain the positions from the diffraction pattern, which can be found by:

\begin{bgreading}
\begin{equation}
\rho(x,y,z) = \frac{1}{V}\sum_{h,k,l}|F(h,k,l)|\exp(i\alpha_{hkl})\exp(-2\pi i(hx,ky,zl))
\end{equation}
\end{bgreading}

The outcome of this reverse Fourier transform is $\rho$, the (electron) density that we want to observe. The three integers h, k, l are known as Miller indices and are an established notation system in crystallography to describe the planes of the different crystal lattices. Input is the structure factor $F$, i.e.\@ the amplitudes that we get from the reflections (spots) in the diffraction data. Through this Fourier relationship, only the spots of the diffraction pattern are required. This means that experimentally, the amplitudes of the diffraction pattern (strengths of the spots arising from constructive interferences) are directly accessible. Unfortunately, the experiment does not provide any information on the associated phase $\alpha$ of the diffraction (contained in the imaginary component $\exp(i\alpha_{hkl})$). Without the phase information, it is not possible to reconstruct the electron density of the crystal cell. This is known as the ``Phase Problem'' in X-ray crystallography. For a more detailed description of the Fourier synthesis and the phase problem see the recommended further readings and \citet{Cowtan2003}. The \panelref{ChDetVal:phases} illustrates this problem using photographs of the two pioneers in (protein) crystallography.

\begin{bgreading}[There is a lot of information in the phases!]
\label{panel:ChDetVal:phases}
The examples in the figure below show how important the phases, $\phi$, are for the reverse Fourier transform (The pictures show two pioneers of X-ray crystallography: Karle on the left and Hauptman on the right). The pictures were forward Fourier transformed (data not shown) which results in the phase and amplitude, and then subsequently reverse Fourier transformed. However, for the reverse transformation, the phases between the two datasets are swapped, i.e.\@ we get Karle with Hauptman's phases, and Hauptman with Karle's phases. This is a rather extreme example, and when the model and real structure are closer (see below) the effect is a lot less severe. Still, clearly, a lot of the information is contained in the phase and not in the amplitudes. This effect is known as ``the phase problem'' in crystallography \cite{Cowtan2003}.

\centerline{\includegraphics[width=0.6\linewidth]{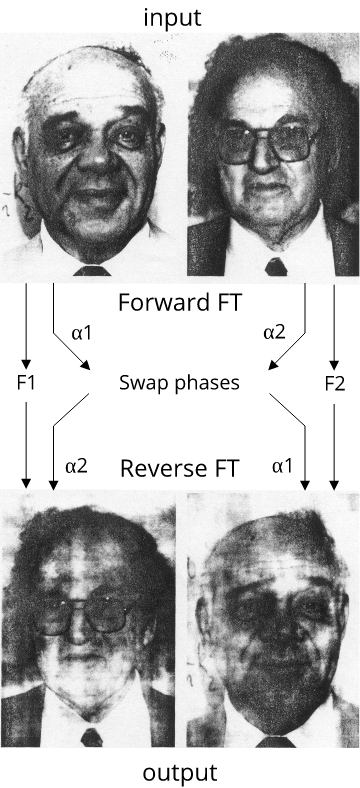}}

There are a number of ways to solve the phase problem. If the resolution of the data is very high (better than 1{\Angs}) and the protein is (very) small, there may be enough information in the amplitudes. Otherwise (in almost all cases for proteins), one can get some of the phase information by giving some of the radiation an ``offset''; the amplitude changes caused by the offset are a measure of the phase. Incorporating heavy atoms into the protein structure does just that, but this requires chemical modification of the protein and carefully replication of the X-ray data collection. Something similar can be done by using not a single X-ray wavelength (which gives cleaner data), but multiple wavelengths. The differences in diffraction of different wavelengths also yield some phase information when compared.

A very different solution makes use of the fact that once you have a reasonable estimate of the structure of the protein of interest, one can calculate the phases from a (forward) Fourier transform of the electron densities derived from the estimated structure. These calculated phases can then be used in the reverse Fourier transform, yielding electron densities. From these electron densities one obtains a new (usually better) set of coordinates for the structure of the protein, which is then used to calculate update phases. This process is usually iterated till convergence. However, convergence is not strictly guaranteed. This approach is sometimes called `molecular replacement'.

\noindent\textsl{Images used with permission from Randy J.\@ Read \cite{Read1997}}
\end{bgreading}

\begin{bgreading} [Electron Diffraction]
\label{panel:ChDetVal:edif}
\centerline{
(a)\hspace{-2em}\includegraphics[height=0.33\linewidth]{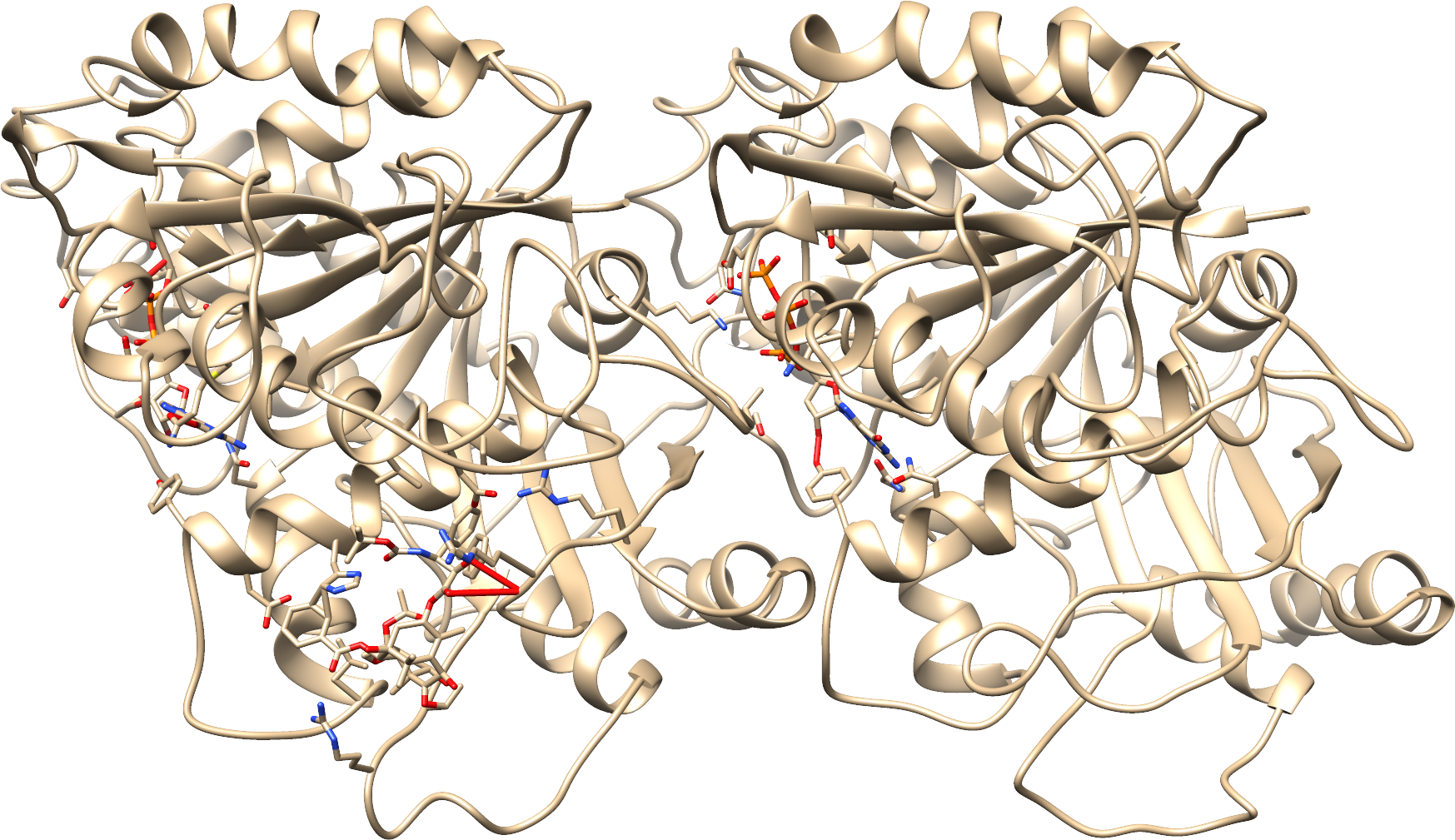}\hspace{-1em}
(b)\includegraphics[height=0.33\linewidth]{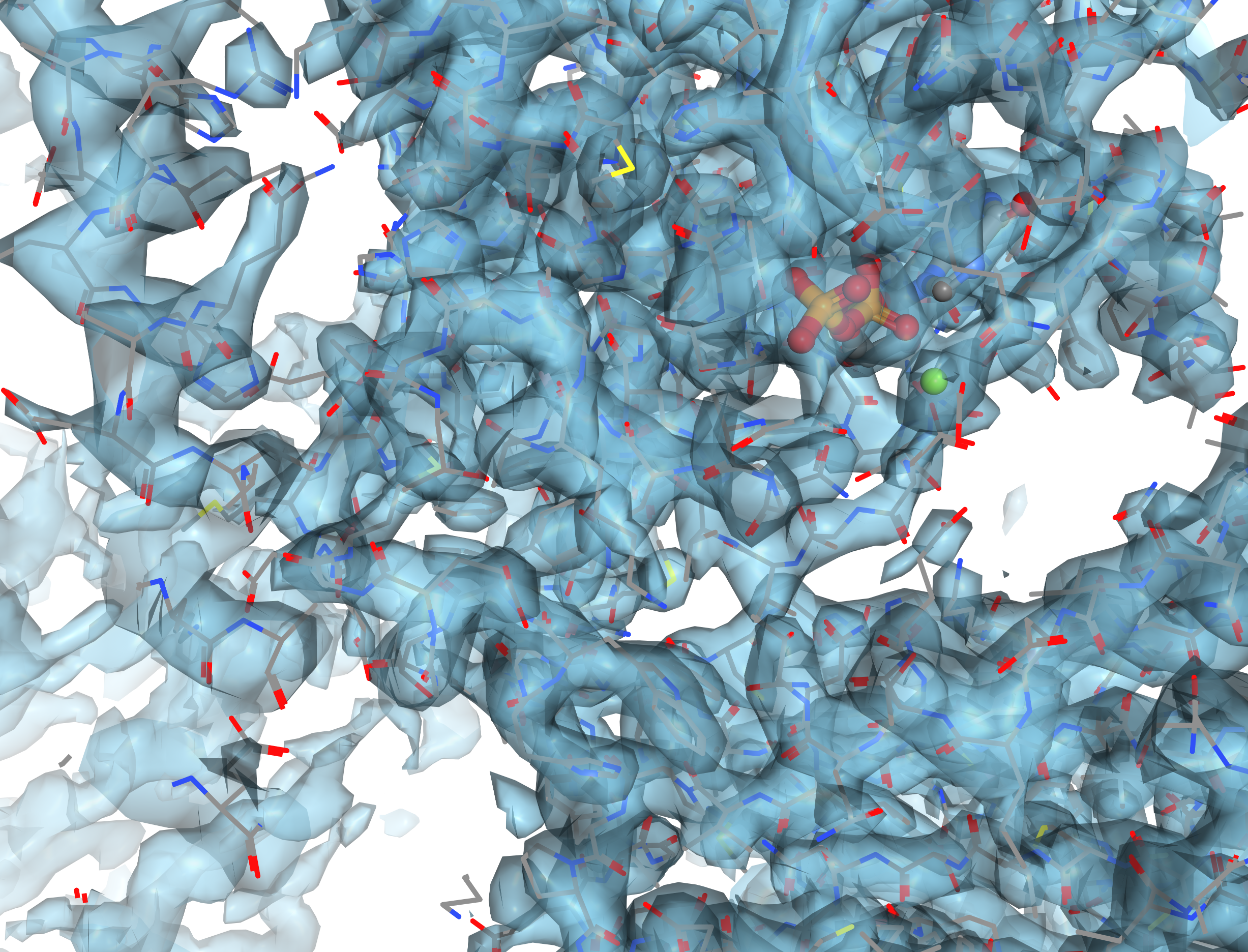}
}
(a) Near atomic resolution structure by electron crystallography and diffraction of the $\alpha\beta$ tubulin dimer by electron diffraction from \pdbref{1tub} \cite{Nogales1998}. Great advantage of electron diffraction over X-ray diffraction, is that electron detectors do allow the direct measurement of phases. But resolution is limited to `near atomic' (3.7 {\AA}), just good enough to identify all secondary structure elements. Most importantly, because of the measurement of phases, the densities are well enough defined to accurately thread the protein chain through them. (b) For comparison the density map of beta-tubulin at 3 {\AA} from \pdbref{5yls} \cite{Yang2018}.
\end{bgreading}

\section{Nuclear magnetic resonance}

Nuclear Magnetic Resonance (NMR) relies on a property of some atomic nuclei known as `nuclear spin' to characterize the structure and dynamics of molecules. The nuclear spin can be thought of as a tiny bar magnet. Fortunately for (bio)molecular scientists, the hydrogen nucleus is magnetic. To reveal its magnetism a very strong magnetic field is needed. When placed in the external magnetic field, the nuclear spin will tend to align with this field. This so-called `up' orientation is the energetically most favourable state. The spin can also transition to a higher energy state, the `down' state, with exactly opposite orientation. The energy difference between these two states corresponds to wavelengths in the radio frequency range and is measured in an NMR experiment. Next to hydrogen, some isotopes of other atoms found in proteins, such as $^1$$^5$N, $^1$$^3$C, are also NMR active and can be used to obtain additional information on the structure. 

\begin{figure}
\centerline{
\includegraphics[width=1.1\linewidth]{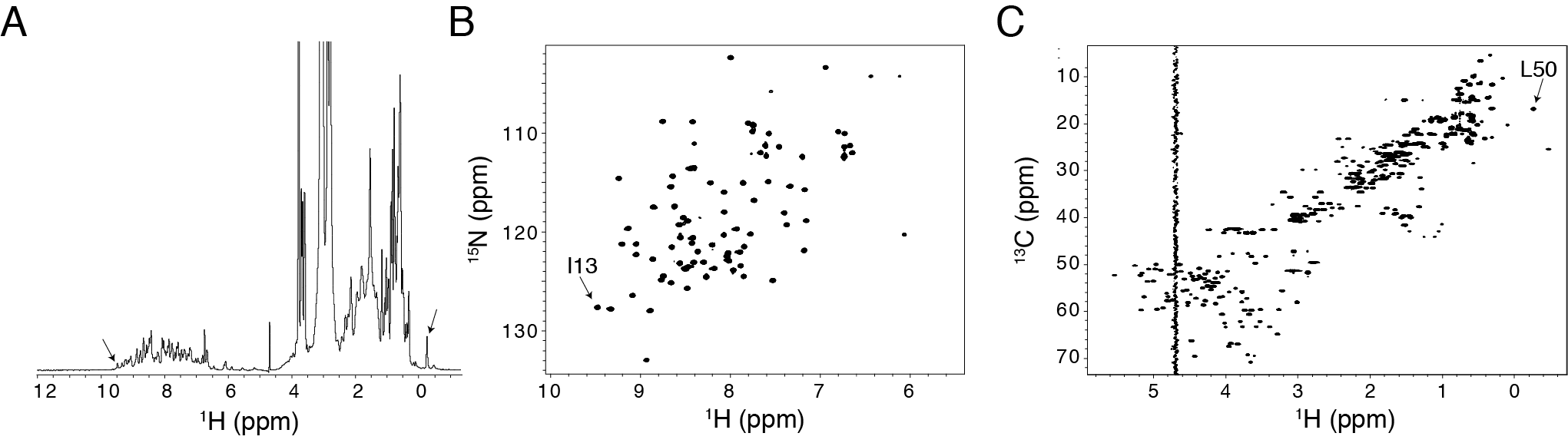}}
\caption{NMR spectra of ubiquitin (76 residues): (a) 1-dimensional hydrogen spectrum. Arrows correspond to the proton signals of the labelled peaks in (b, c). Intense peaks between 3 and 4 ppm are from the buffer. (b) 2-dimensional hydrogen-nitrogen (HN) spectrum. The backbone NH signal of Ile13 is labelled. (c) 2-dimensional hydrogen-carbon (HC) spectrum. One of the methyl CH$_3$ signals of Leu50 is labelled. The vertical ridge is from the water signal. The spectrum axes (horizontal in a, and both in b,c) are expressed in parts-per-million (`ppm') deviation of the frequency with respect to a standard reference. Due to two spectrum dimensions being used in the 2D experiments, most of the overlapping peaks that appear in the 1D spectrum are resolved. The HN spectrum shows signals of the backbone amide NH groups and signals from NH group in side chains of some amino acids. This spectrum is very sensitive to changes in protein conformation, see panel `NMR-based modelling of protein complexes'. The HC spectrum shows signals of the -CH, -CH$_2$ and -CH$_3$ groups in aliphatic side chains as well as the backbone CH group at the alpha-position. 
}
\label{fig:ChStrucDet-nmr-1d-2d}
\end{figure}

The frequency associated with transitions between the two levels not only depends on the type of atom (say hydrogen vs.\@ phosphorus) but also on the local chemical environment of the atom (say a hydrogen atom in a methyl-group vs.\@ in an aromatic ring). The sensitivity of the transition frequency to the local environment of the spin is the basis for the application of NMR in chemistry. To stress this importance, the transition frequency is usually called the `\emph{chemical shift}'. Chemical shifts of a given nucleus are very small and therefore expressed in parts per million (ppm), based on the relative change in transition frequency compared to a standard compound (usually tetramethylsilane, which is added as a reference in the experiment). The NMR spectrum of a protein will typically contain hundreds to thousands distinct signals, because each hydrogen nuclear spin will have a slightly different chemical environment. This results in a very crowded spectrum, as can be seen in \figref[A]{ChStrucDet-nmr-1d-2d}, showing the NMR spectrum of ubiquitin. To use NMR in an intelligible manner two things have to be accomplished: first, the different signals need to be resolved; second, structural information about the relative position of the nuclear spins needs to be encoded in the NMR signal.

To resolve the overlapping signals, one usually performs a two-dimensional or three-dimensional experiment. Here, the chemical shifts of two or three different nuclear spins are measured simultaneously along the different dimensions of the spectrum. Typically, the additional dimensions are used to measure the chemical shift of the heavier isotope of nitrogen and carbon ($^1$$^5$N and $^1$$^3$C) which also have spin. \figref[B and C]{ChStrucDet-nmr-1d-2d} show the 2D H-N and H-C spectra of ubiquitin, illustrating that the addition of another chemical shift dimension allows to resolve nearly all signals. The \panelref{ChDetVal:NMRcosy} shows some more detail in part of a 2D spectrum of a small peptide. The principle for generating 2D spectra can be extended to include multiple dimensions. In practice, 3D NMR, e.g.\@ H-N-C, spectra are common in protein structure determination and are crucial to find out which peak corresponds to which atom (see the panel \panelref{ChDetVal:NMRshifts}).

To encode structural information in the NMR signal, again multi-dimensional experiments are used. The spectra in \figref[B and C]{ChStrucDet-nmr-1d-2d} already encode information about the secondary structure, as the chemical shifts of the backbone nuclei are sensitive to the backbone dihedral angles. To get data on the 3D fold of the protein, one makes use of the fact that the nuclear spin can `sense' the presence of other nearby nuclear spins through their mutual magnetic interaction. This allows one to transfer the magnetic energy of one nuclear spin to another spin, this is known as the nuclear Overhouser effect (NOE). In a dedicated 2D experiment one then measures the chemical shifts of the two spins involved. Since the magnetic interaction between spins is distance dependent, the energy transfer and thus signal intensity (`NOE intensity') is also distance dependent. In this way, the distances between nuclei can be measured. For more explanation, please refer to \panelref{ChDetVal:NMRshifts}.

\begin{figure}
(a) \includegraphics[width=0.45\linewidth]{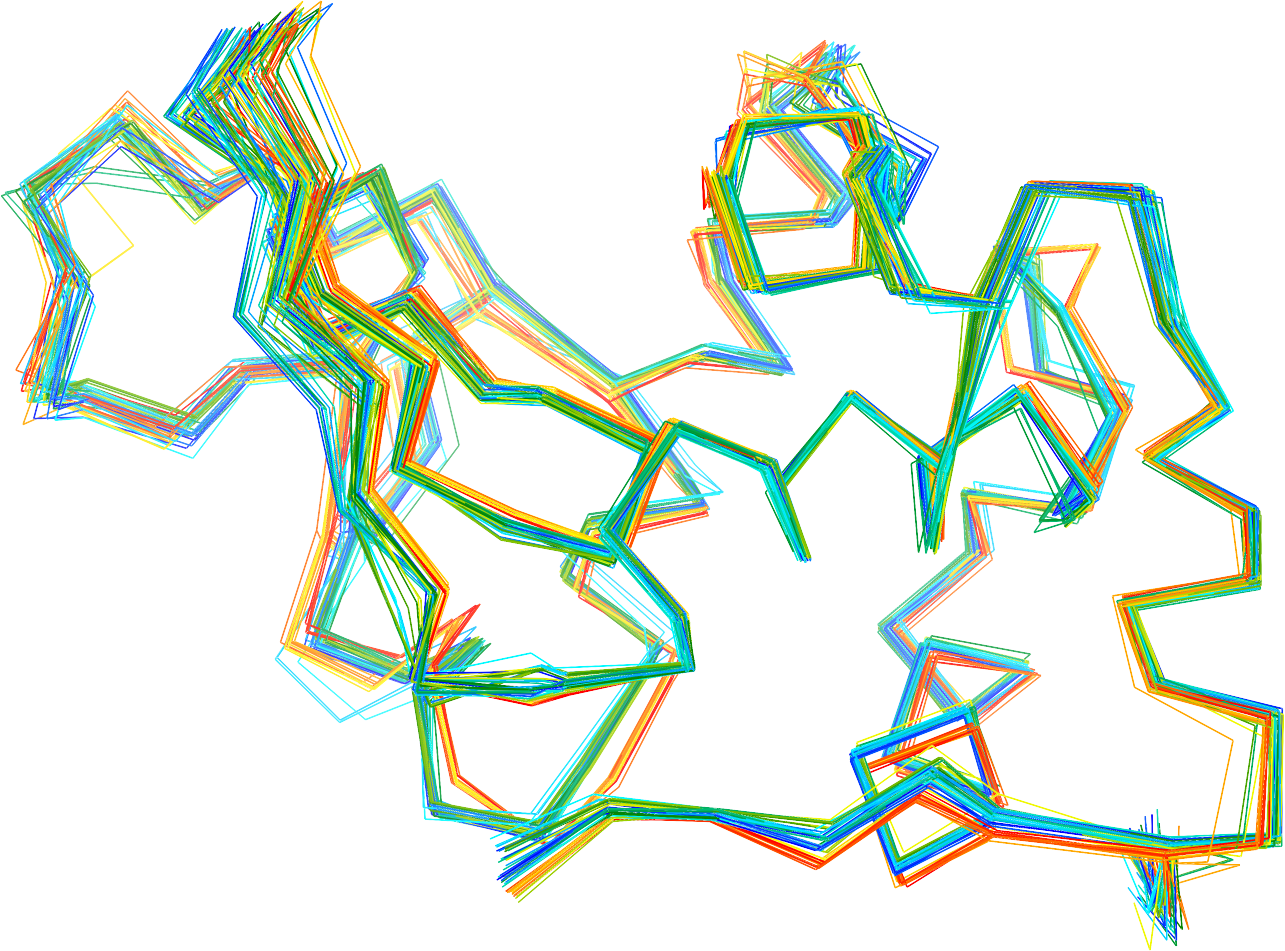}
(b) \includegraphics[width=0.45\linewidth]{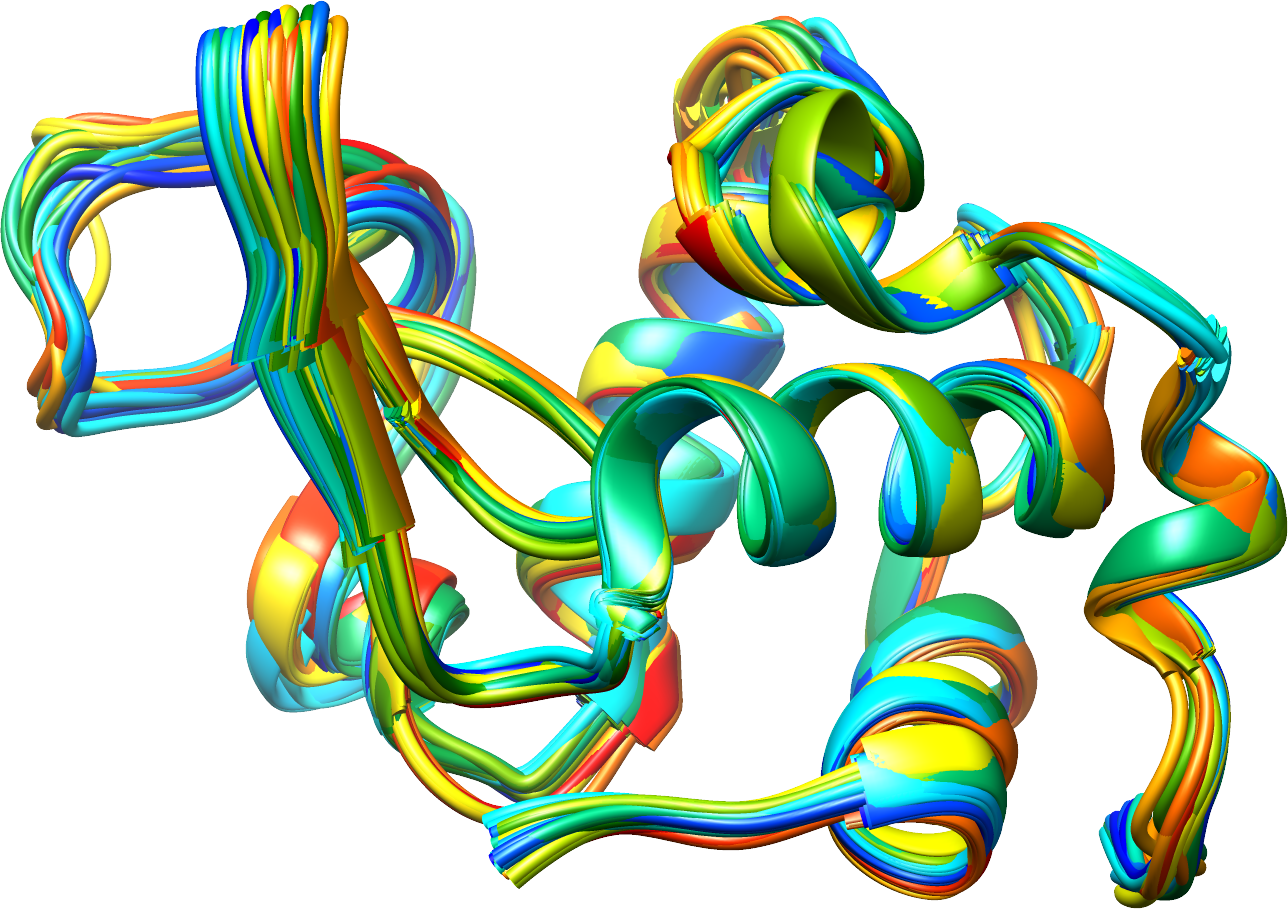}
\caption{Representation of the ensemble of NMR solution structures by (a) an explicit ensemble of backbone traces, and (b) an overlay of `cartoon' renderings. One can clearly see that variability between conformations is different in different places of the protein. \pdbref{1e8l} \cite{Schwalbe2001}}
\label{fig:ChStrucDet-NMR-ensemble}
\end{figure}

\begin{bgreading}[NMR two-dimensional spectrum]
\label{panel:ChDetVal:NMRcosy}

In the Correlation Spectroscopy (COSY) experiment, the setup is such that energy transfer is predominant through chemical bonds. The figure shows an example for a small peptide \cite{Feenstra2002}. Diagonal peaks are where the absorbing hydrogen nucleus also emits, and cross peaks correspond to energy transfer; note that transfer here is induced via chemical bonds. The cross-peak connections between the $\alpha$ and $\beta$ hydrogens and between $\beta$ and $\gamma$ hydrogens for the Valine and Threonine are traced out. This allows us to identify frequencies with unique individual protons in the molecule, which is rather important, because we cannot know beforehand which hydrogen will respond to which frequency. Since we know the sequence of our molecule, using the 2D-COSY spectrum we can trace out which hydrogen atom is where in the spectrum. The process is called `chemical shift assignment': the assignment of which specific frequencies correspond to which atoms.

\centerline{\includegraphics[width=0.6\linewidth]{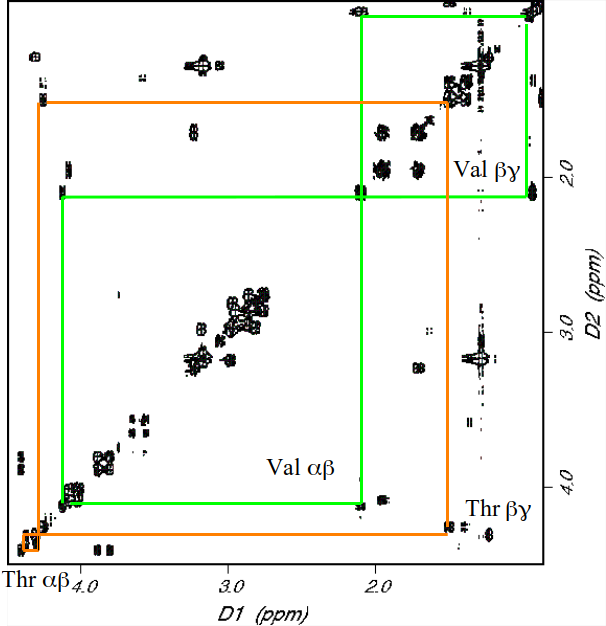}}

\end{bgreading}

The distances and dihedral angles that are measured from NMR spectra are combined with a priori knowledge of the amino acid sequence of the protein and the structures of the amino acids to derive the 3D protein structure. It should be noted that NMR distance and angle data do not give direct access to the atomic coordinates, but these rather serve as constraints in the structure calculation process. The structure is fitted to these constraints in a calculation much like a molecular dynamics simulation or Monte Carlo sampling (see \chref[n]{ChMD} and \chref[n]{ChMC} for more on those techniques). Usually multiple, slightly different solutions are possible, which results in the typical bundle appearance of NMR structures, see \figref{ChStrucDet-NMR-ensemble}. Some of the variability in the models may show real structural fluctuations in solution, where other variation may indicate a lack of data or accuracy. 

This procedure for structure determination by NMR works well for protein structures up to about 300 residues. Importantly, proteins need to be isotope-labelled with $^1$$^5$N and $^1$$^3$C, which works best if the protein can be expressed and purified from E.~coli. Samples are typically solutions of the protein of interest, but can also be semi-solid samples of membrane proteins embedded in native membranes or suitable membrane-mimics. In both cases the crystallisation step is not required. Next to structure determination, important applications of NMR are the study of intrinsically disordered proteins, the study of protein dynamics and the study of protein-protein interactions (see the \panelref{ChDetVal:NMRppi})

\begin{bgreading}[NMR chemical shift assignments and structure determination]
\label{panel:ChDetVal:NMRshifts}
\centerline{\includegraphics[width=\linewidth]{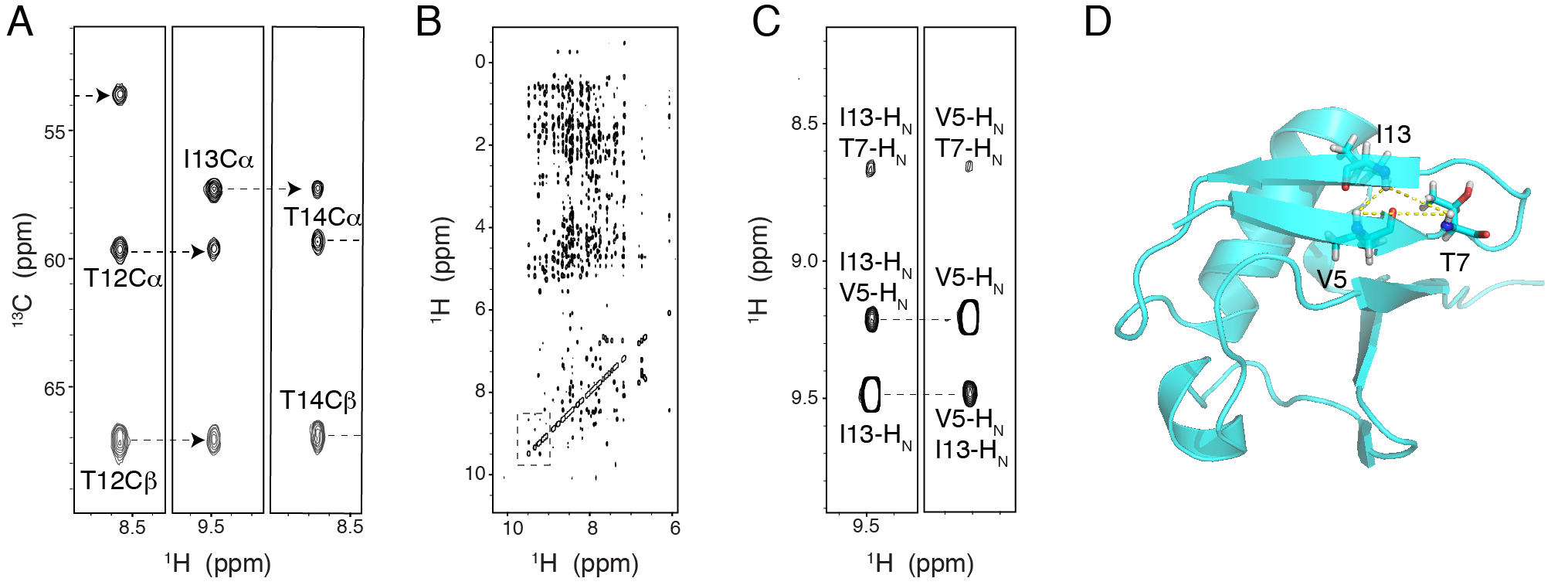}}

One of the main tasks in protein studies by NMR is the assignment of the signals for the backbone nuclei, in particular of the backbone amide NH groups. In this process, the transition frequency or chemical shift of a certain signal is assigned to a particular atom in the protein. This is achieved through the analysis of a dedicated set of so-called triple resonance NMR experiments. In these experiments, the chemical shifts of the amide NH and backbone $^1$$^3$C of each residue are measured. For each residue also the chemical shifts of the previous residue are measured. The resulting puzzle is to trace back the amino acid sequence by finding the matching connections from one residue to the next. This is illustrated in panel A for residue I13. Panel B shows a segment from a so-called NOESY experiment that is used to measure the distance between atoms. Each signal corresponds to an energy-transfer between nearby spins. The signal intensity can be used to derive the interatomic distance, where intense peaks mean that the atoms are close ($<$ 3.5~{\AA}) and weak peaks that the atoms are within 5~{\AA}. When atoms are further than 5 to 6~{\AA} apart, the magnetic interaction between spins is too weak to result in transfer and no peak will be observed. The expanded plot in panel C shows NOE peaks observed for I13. Each peak corresponds to two chemical shifts, one along the horizontal and one along the vertical dimension, each corresponds in turn to a particular atom, as shown in the labels. The intense peak labelled ``I13-HN V5-HN'' is the result of transfer between amide protons of I13 and V5, residues that are far apart in the primary sequence. This means that these two are in close proximity in the 3D fold of the protein, as shown in panel D. Such information is particularly valuable when determining structures.
\end{bgreading}


\begin{bgreading}[NMR-based modelling of protein complexes]
\label{panel:ChDetVal:NMRppi}
\centerline{\includegraphics[width=\linewidth]{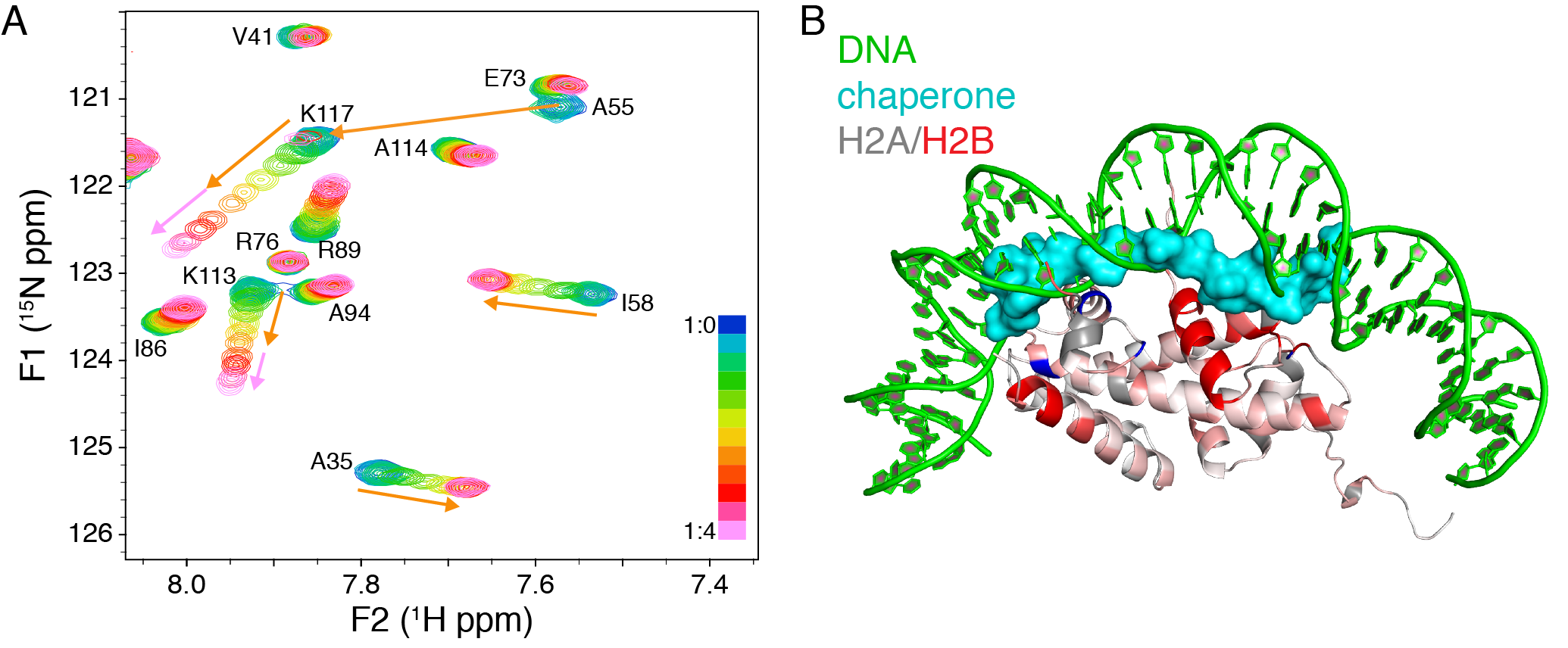}}
The position and intensity of peaks in the 2D NH spectrum such as the one shown in \figref[B]{ChStrucDet-nmr-1d-2d} is very sensitive to the precise conformation of a protein, it is like a protein's fingerprint. It is therefore very suitable to investigate the interaction of a protein with a binding partner. The partner can be anything ranging from an ion, a small molecule ligand, peptide or an entire protein complex. Addition of the binding partner to the protein will result in changes in the local environment of nuclei that are part of the binding interface. Thus, the chemical shift of these nuclei will change and the corresponding peak will appear at different position in the NMR spectrum. By examining which peaks are perturbed, the binding interface on the surface of the protein can be mapped. Quantitative analysis of the changes can be used to extract binding affinity and the life time of the complex. Panel A shows such an experiment to find out where a histone chaperone binds on the surface of the histone H2A/H2B dimer \cite{Corbeski2018}. The spectrum is the NH spectrum of a histone dimer in which only H2B is isotope labelled. Each signal thus corresponds to a backbone amide NH group in H2B. Ones that are in the interface show a gradual change from their free state position (blue) to their bound state position (magenta). These changes were used to derive the dissociation constant and life time of the complex. Using the mapped interface, the structure of the complex could be modelled using protein docking software. Panel B shows how the chaperone binds to a surface of the histone dimer and how such binding is able to prevent DNA binding. This approach of mapping binding interfaces and modelling of protein-protein complexes is applicable also to very large protein complexes such as the nucleosome \cite{Kato2011}.
\end{bgreading}

\section{Cryo electron microscopy (cryo-EM)}
\label{sec:ChDetVal:cryoEM}

Next to X-ray and NMR, cryo electron microscopy (cryo-EM) has become more and more popular in the past decade to obtain structural information on protein and biomolecular systems. With cryo-EM, as it is a microscopy technique, we can directly observe the objects of interest. The fundamental limitation of this technique is the wavelength of the electrons. In principle they can be tuned to any desired wavelength, however shorter wavelengths are progressively higher in energy. At some point the energy input into the protein will quickly destroy the sample. Atomic resolution is thus only obtainable in certain specific conditions, including cooling to extremely low temperatures.
The method allows to image two-dimensional crystals that can be obtained from membrane-bound proteins, which are typically hard to crystallize into `normal' 3D crystals for X-ray diffraction (see also \panelref{ChDetVal:challenging}). Especially large-scale structures such as kinesin, tubulin, $\beta$-amyloid fibres, and virus capsids have become routine work for the cryo-electron microscopist. Unfortunately, unlike optical microscopy, EM does not work on `live' samples. Because of the wavelength/energy problem mentioned above, it is hard to get sufficient contrast at high resolutions without damaging the sample. 
In cryo-EM, the sample is cooled with a cryogen, liquid nitrogen or helium, allowing the electron density of the protein molecules to be observed without the `blurring' that is caused by atomic motions at room temperature. To obtain atomic-level resolution, cryo-EM density maps are often combined with X-ray structures in what can be thought of as an `X-ray jigsaw' solution.

\begin{bgreading}[Single Particle Electron 3D reconstruction]
\label{panel:ChDetVal:1pem}
\centerline{
(a)\includegraphics[width=0.45\linewidth]{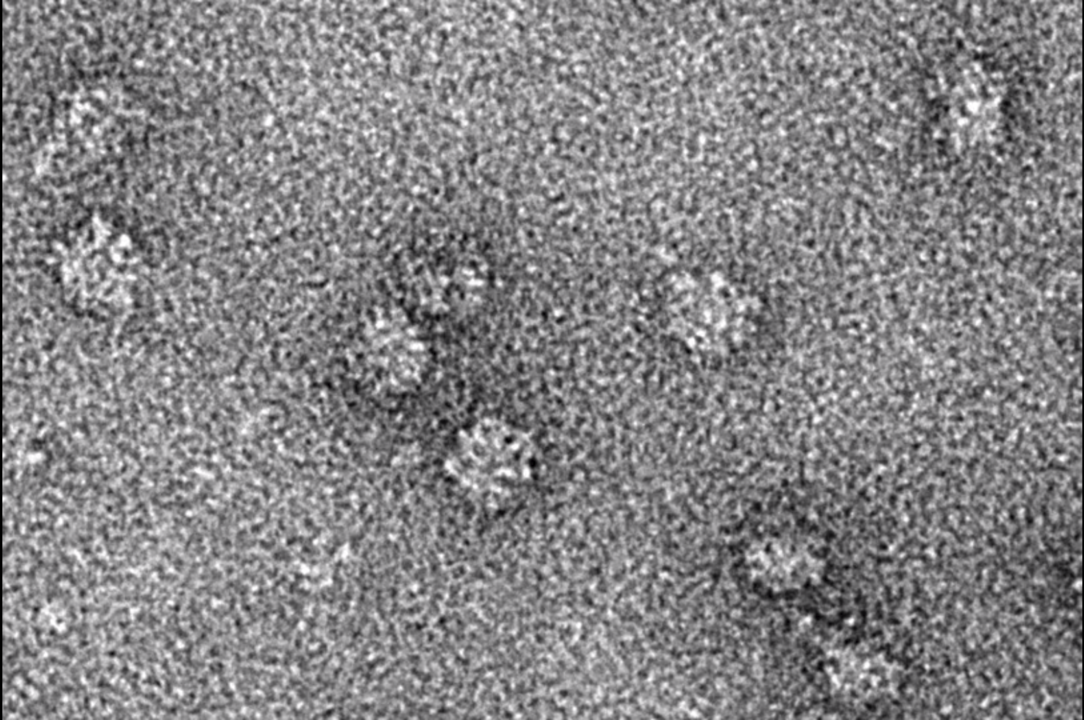}
(b)\includegraphics[width=0.45\linewidth]{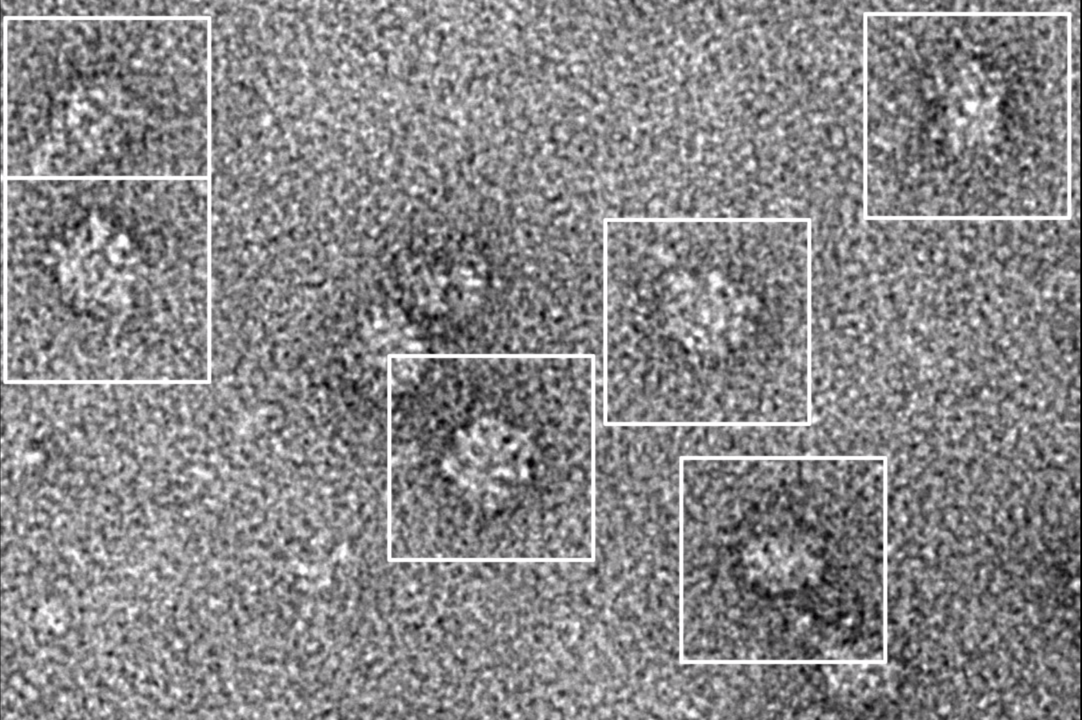}
}
\centerline{
(c)\includegraphics[width=0.45\linewidth]{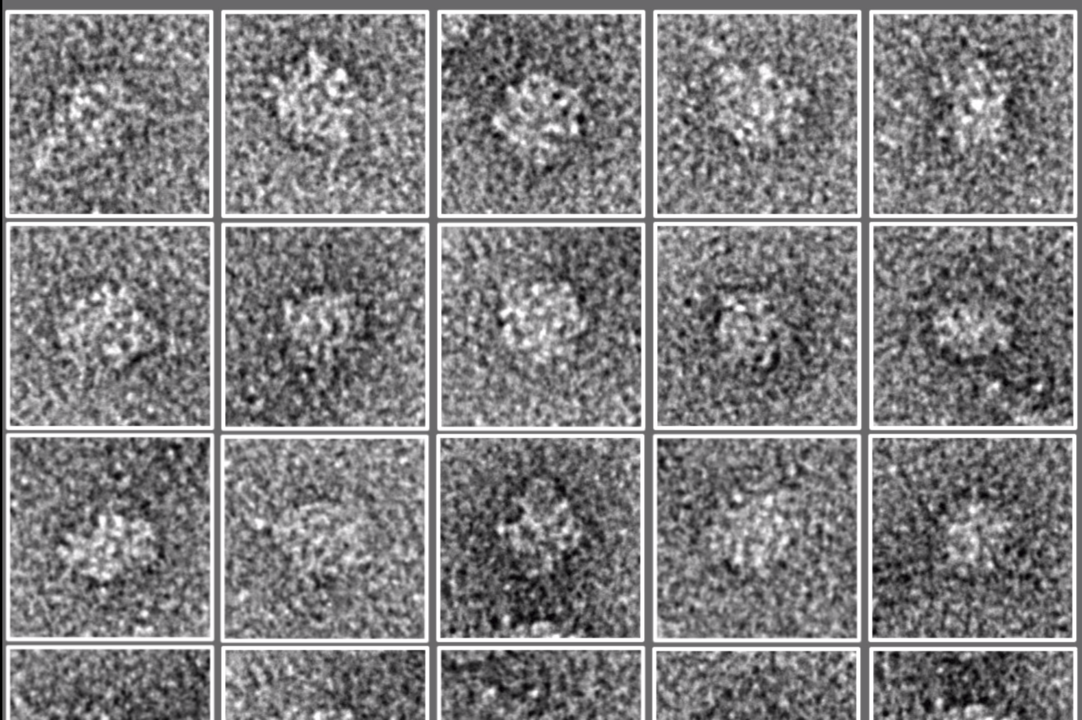}
(d)\includegraphics[width=0.45\linewidth]{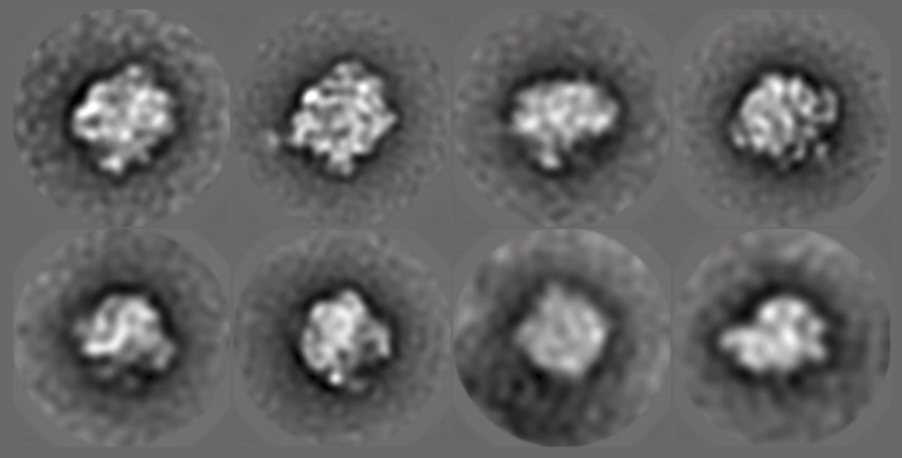}
}
Electron microscopy allows single particles to be imaged, but to obtain more detail (higher resolution), shorter wavelengths are needed which means a higher energy beam. This greatly increases the damage done during data acquisition. To solve this, an average over many such particles can be used. Importantly, these particles do not need to be in a crystal, and that can be a huge advantage. 

Images are taken of individual particles, such as virus capsid structures, lying on a grid (a). To minimize radiation damage the exposure is very brief and the contrast between the image and the background is minimal.  In this method particles are selected (b) and sorted into different classes (c), which correspond (approximately) to the orientations in which the particle lies on the grid.

Each class (orientation) is then averaged to improve the signal to noise (d), which are then used to reconstruct a 3D image (not shown here). The method depends on the molecule falling on the grids in a fairly random way. Molecules that preferentially fall in only a few orientations need to be treated specially by tilting the grid. 

Images courtesy of Peter Shen and Janet Iwasa (University of Utah) \url{https://cryoem101.org/chapter-1/}.
\end{bgreading}

\begin{figure}
\centerline{
(a)\includegraphics[height=0.65\linewidth]{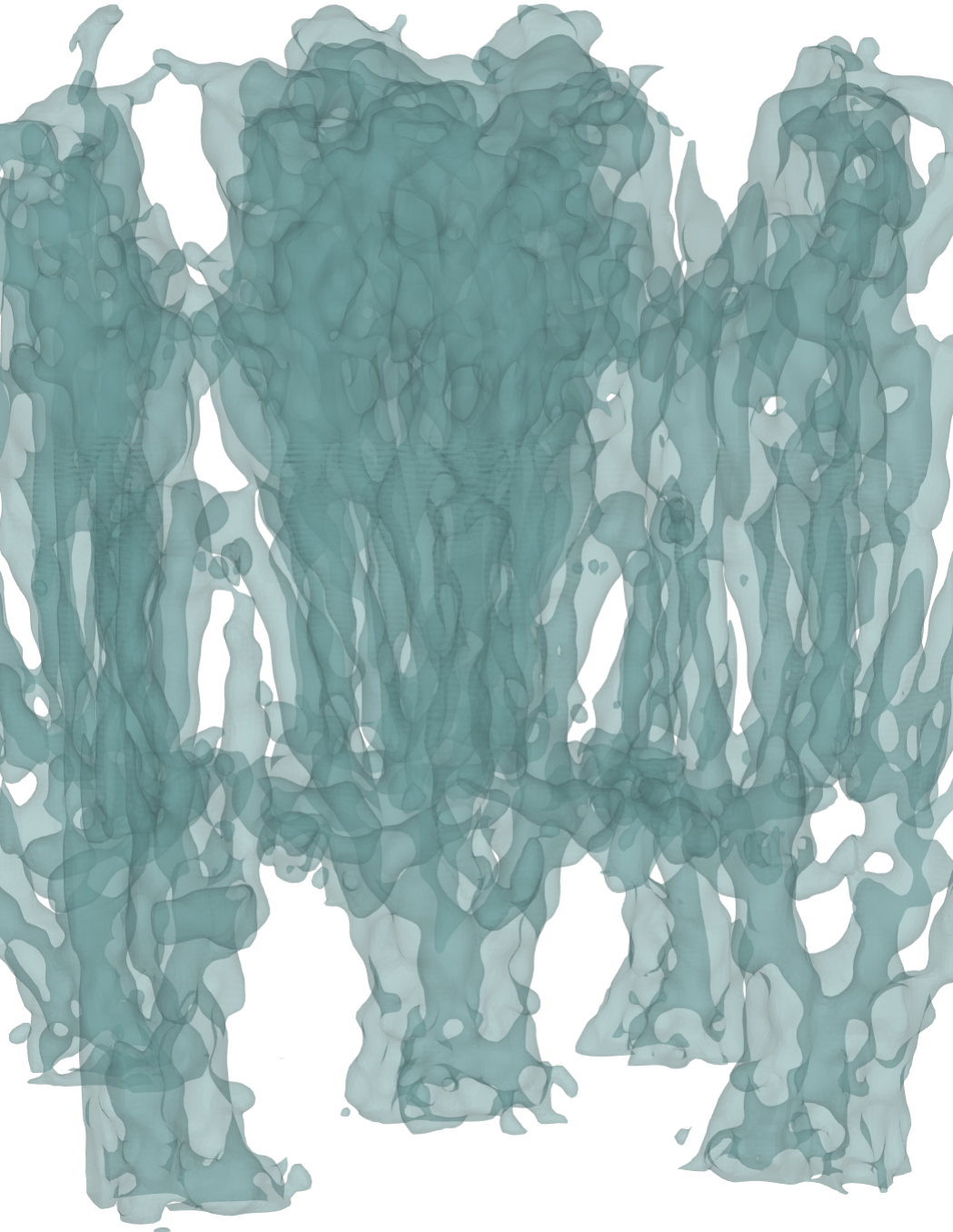}
(b)\hspace{-1em}\includegraphics[height=0.65\linewidth]{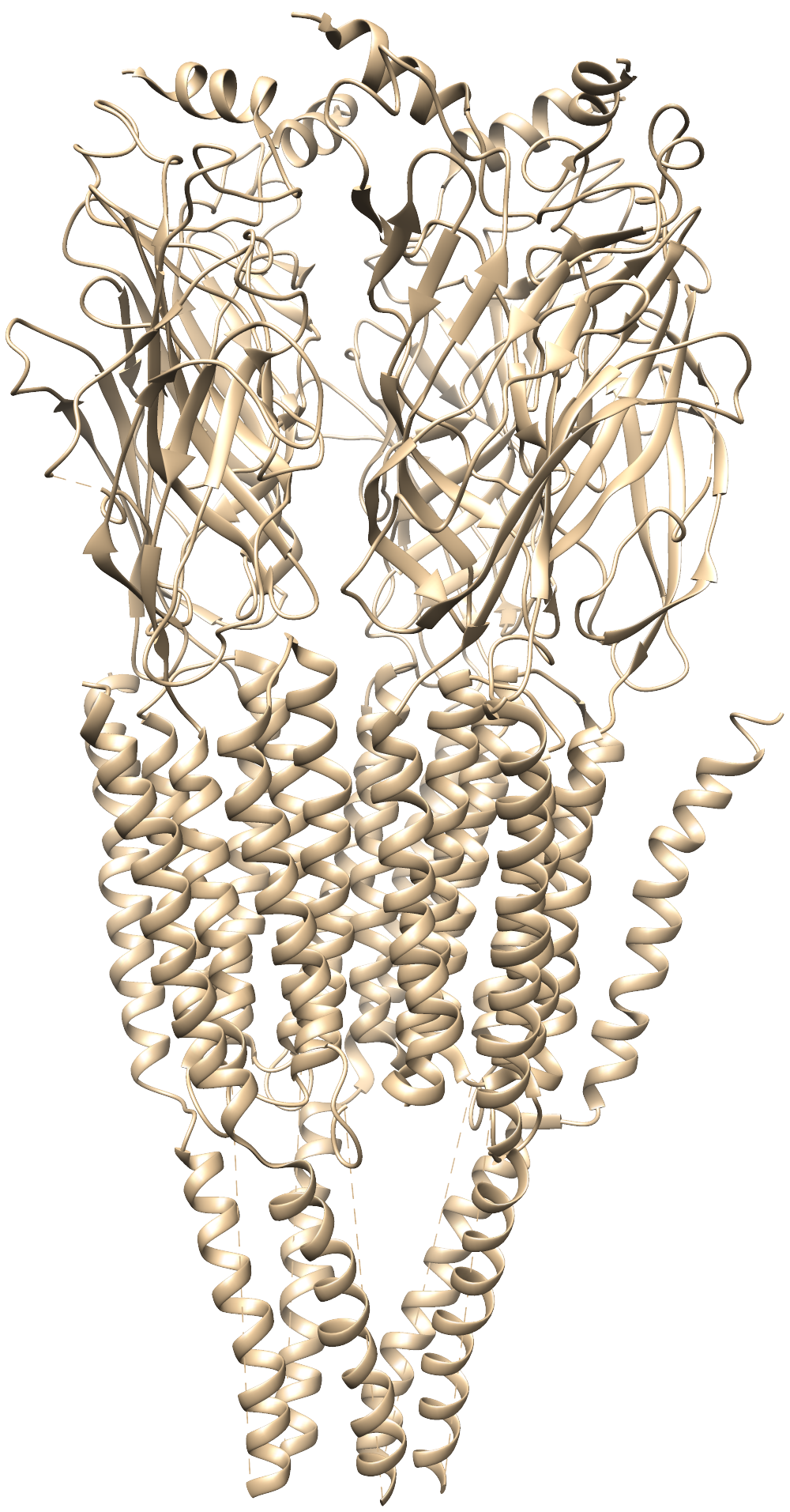}\hspace{-1em}
}
\caption{The first low resolution density map of a membrane bound receptor, the nicotinic acetylcholine receptor, was created at 9{\AA} \cite{Unwin1993}. Later, greatly improved resolution of the cryo-EM experiments yielded a maps at 4{\AA}, allowing atomic models to be constructed \cite{Unwin2005,Unwin2012}. (a) Overview of the density map \pdbref{4aq9}. (c) Full details of the protein structure \pdbref{2bg9}. Images generated by LiteMol \cite{Sehnal2017}.}
\label{fig:ChStrucDet-nac}
\end{figure}

\figref{ChStrucDet-nac} shows the membrane bound receptor for which the first density map was measured in 1993, at low resolution (9{\AA}) by \citet{Unwin1993}. Those maps did not allow the modelling of atomic coordinates. Given the challenges of crystallography for these type of proteins, even this low resolution can give valuable insights. Later, the same protein was solved in the same lab at much improved resolution (4{\AA}), where secondary structure elements may be clearly resolved, but not all atomic details become available \cite{Unwin2005,Unwin2012}.

\begin{figure}[h]
\centerline{
\includegraphics[height=0.8\linewidth]{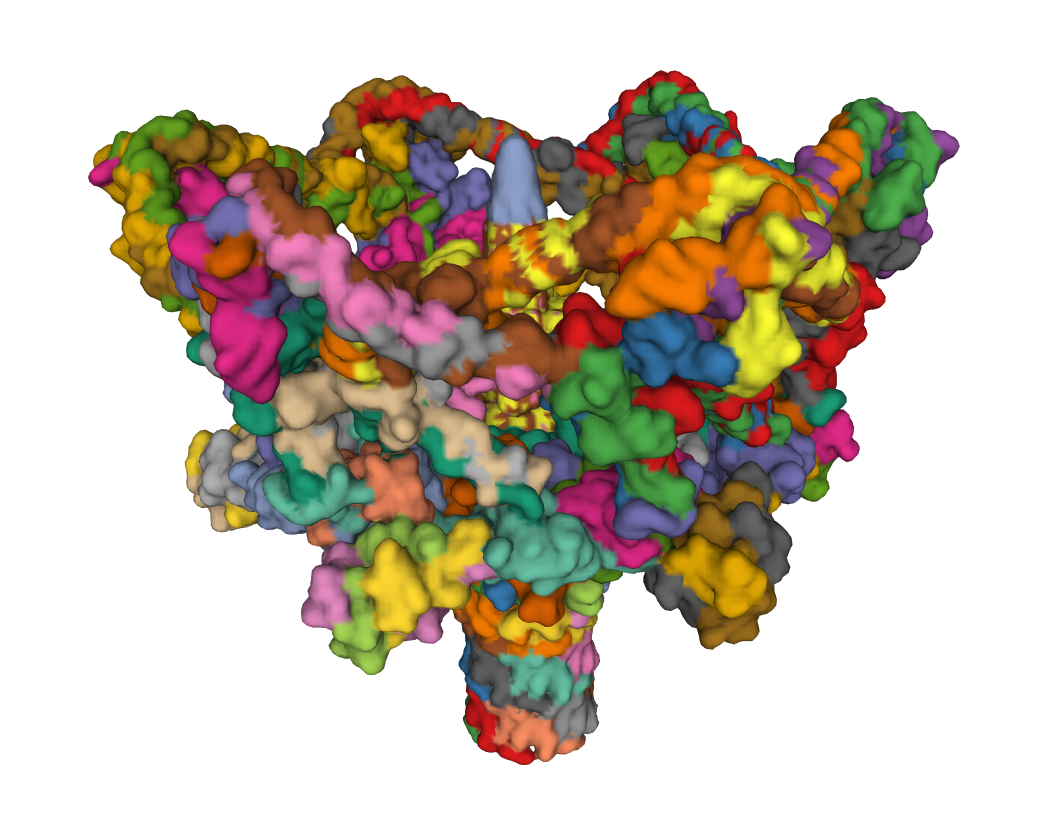}
}
\caption{The complex is the tail of a bacteriophage (virus that infects bacteria); the tail contracts to insert the DNA into the host bacterium. It was imaged in EM at 17{\AA} resolution by \citet{Leiman2010}; the image shown here is of the attachment baseplate and tube, also with cryo-EM at 4.1{\AA} \cite{Taylor2016}. The whole complex measures 1200 {\AA} (120 nm) in length and has an atomic weight of 20 million Daltons (one amino acid on average is about 134 Dalton). There are about 20 different proteins present in the complex, most in (very many) multiple copies. Image generated from \pdbref{5IV5} using the PDB viewer \cite{Berman2000}.}
\label{fig:ChStrucDet-T4-tail}
\end{figure}

Methods for single molecule cryo-EM structure determination have been developed since the early nineteen nineties, e.g.\@ on a skeletal muscle calcium channel \cite{Radermacher1994}. See \panelref{ChDetVal:1pem} for a more detailed description of an application to viral envelope structures. Many methodological advances have since pushed the achievable resolution to near atomic \cite{VanHeel2000,Frank2002,Li2013a}, and applications of large macromolecular machines, e.g.\@ \citet{Baumeister2000}. Some landmarks include the calcium channel already mentioned \cite{Radermacher1994}, the E.~coli ribosome \cite{Malhotra1998}, icosahedral viruses \cite{Baker1999}, and the plant photosystem II complex \cite{Barber2000}.
A particularly impressive example of `cryo-EM reconstruction', or X-ray/EM jigsaw solution of the structure of a large complex is the structure of the tail of bacteriophage T4, shown in \figref{ChStrucDet-T4-tail} \cite{Leiman2010}. Comparing the structure in the relaxed and contracted state (not shown here) of the tail helps understanding the infectious mechanism.

However, these are just some examples on the applications of cryo-EM. The field has been developing rapidly over the past decade and the new experimental advances allow to address more and more challenging structures and molecular targets. If you are interested to know more about the possibilities, you should have a look at the review by \citet{Cheng2018} on the rise and evolution of cryo-EM. The \panelref{ChDetVal:T3SS} shows one of the latest developments. 

\begin{bgreading}[Type III Secretion System]
\label{panel:ChDetVal:T3SS}
The Type III Secretion System is one of the mechanisms by which pathogens infect human cells. It is anchored with a `basal plate' in the bacterial inner and outer membrane, and extends a `needle' filament towards the target (human) cell.

\centerline{
(a)\includegraphics[width=0.8\linewidth]{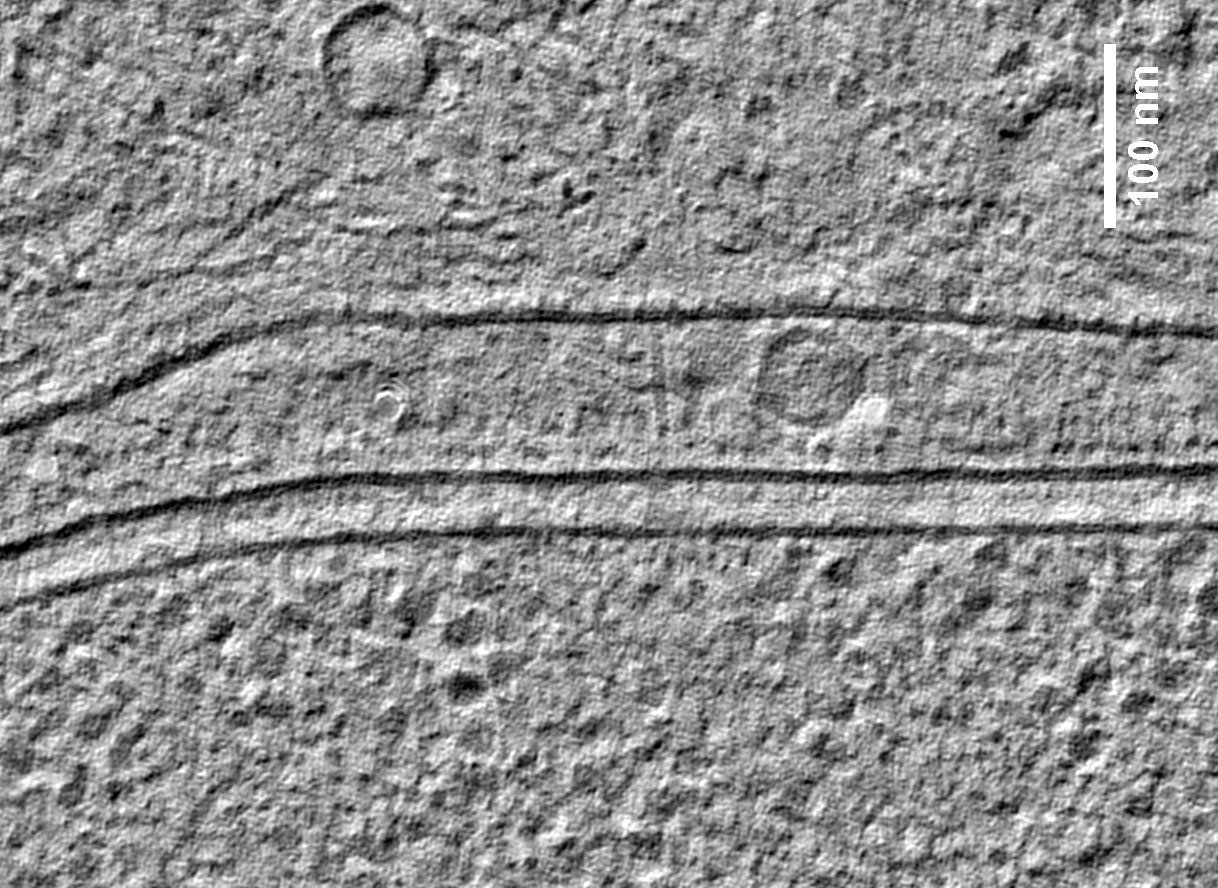}
}
\centerline{
(b)\includegraphics[height=0.4\linewidth]{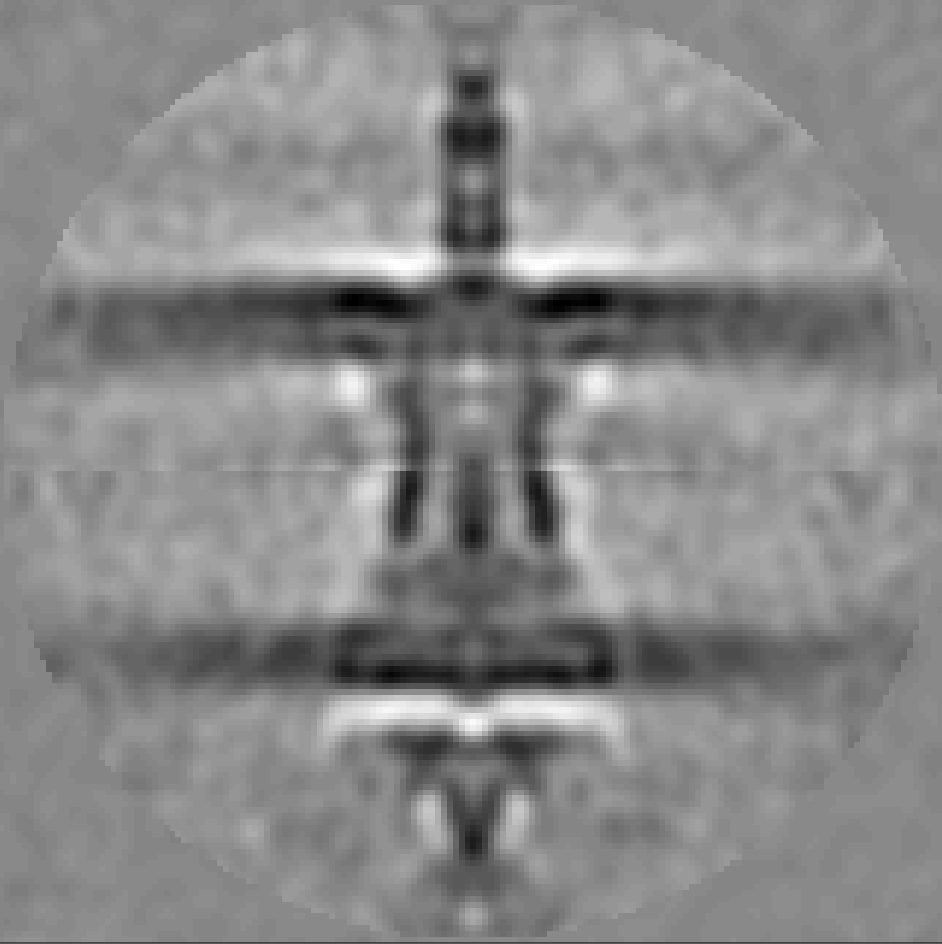}
(c)\hspace{-0.5em}\includegraphics[height=0.4\linewidth]{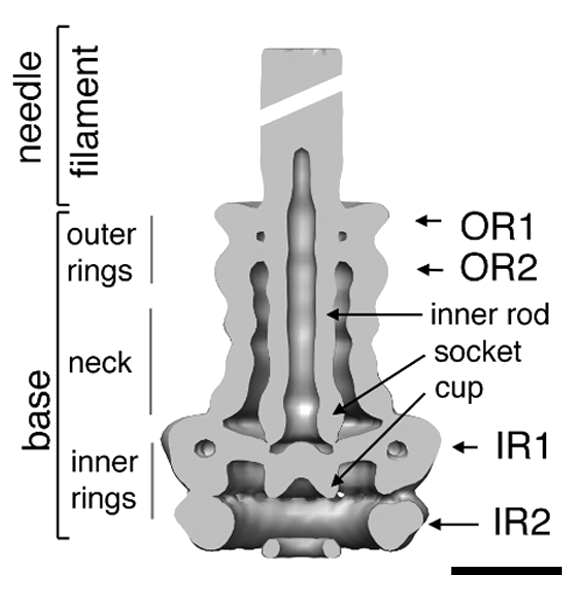}\hspace{-0.5em}
}\noindent
The top image shows a bacterial cell inside a human cell (a); the bottom two dark lines are the bacterial inner (lowest) and outer (top) membranes. In the centre of the image one can see a dim vertical line; this is the needle that extends to the human cell membrane (top horizontal line). Impressively, from a collection of such images, a higher resolution comput 3D image can be generated, which is shown in (b), this allows the complex and its mechanism to be studied in its natural environment inside the cells, and possibly in different conformational states. The layout of the basal complex is shown in (c) \cite{Schraidt2010}. 
{\noindent\footnotesize Images courtesy of Peter Peters, Institute of Nanoscopy (IoN), Maastricht University \url{http://www.maastrichtuniversity.nl/nanoscopy}}
\end{bgreading}

\section{Other structure determination methods}

Besides the ``high-resolution'' structure determination methods we have seen so far, there are many more techniques that allow us to obtain structural information of proteins or relevant biological systems. These are often used in combination with molecular simulations as they do not provide the same resolution as X-ray, NMR, and EM. Small-angle X-ray scattering (SAXS) uses the scattering patterns of X-ray radiation to obtain information on the outside shape of molecules in solution. Several spectroscopic methods such as circular dichroism and infrared spectroscopy can be extremely useful to obtain qualitative structural insight. For example, one may follow conformational changes of proteins upon ligand binding or under different varying conditions, e.g.\@ by increasing/decreasing the pH or by adding salts.

\section{Dynamics and flexibility}

In spite of the rigid-sounding name, protein crystal structures are actually quite dynamic. This also means proteins can retain some (or all) of their biological function, like ligand binding or enzyme turnover. For example, most enzymes are still active in the crystal form, although sometimes much slower than in solution. In a simulation of the native dynamics of a $\beta$-barrel fatty acid binding protein, we can see everything moves all the time but the overall structure of the protein remains the same folded $\beta$-barrel. Also the bound fatty acid molecule remains in its place, even though it wiggles back and forth a lot. Some water molecules that start inside the protein eventually `escape'. But also some that start outside the protein, in the `bulk' water, can find their way into the interior during the course of the simulation.

Some enzymes, however, require large scale motions of the protein structure. Large-scale motions are not possible due to the tight packing of the protein molecules in the constraints of the crystal lattice. In some cases such a motion can be induced for example in enzymatic activity or activation of a receptor by adding the ligand. When a substrate is added to the crystal, and binds to the enzyme or receptor, this binding may then cause the crystal physically break. This also shows how strong some molecular motions can be. 

Even though protein crystal structures give us initially a rigid picture of the protein, the data does contain some information that may also be interpreted in a dynamic sense. The B-factors, or temperature factors as they are also sometimes called, indicate how well the local electron density fits the atoms placed in it. A low B-factor means the fit is very good, a high B-factor means it is poor.
There are two contributions to this goodness of fit. One is variations between the molecules in the crystal. The refinement process of X-ray crystallography assumes all molecules to have identical conformation and orientation throughout the crystal. For proteins this does not strictly hold, and any heterogeneity in the crystal arrangement will result in spreading of the electron density, and hence the diffraction signal. The second contribution is dynamics. Even if, in principle, all protein molecules are identically oriented, there will still be motion going on in them. This leads to a `smearing' of the densities observed (similar to the blurring effect you get when taking a picture of a fast moving object), and hence higher B-factors. This also is the reason that crystal structures are often recorded at low temperature; this slows down the molecular motions and diminishes the `blurring' effect.
From high-resolution low-temperature structures we now know that, typically, the effect of variation or heterogeneity between the protein molecules in the crystal is minor. So, in general it is relatively safe to interpret high B-factors as indicating high mobility in the structure.

\begin{bgreading}[Allosteric motions and time-resolved crystallography]
\label{panel:ChDetVal:allosteric}
(a)\hspace{-1.5em}%
\includegraphics[width=0.5\linewidth]{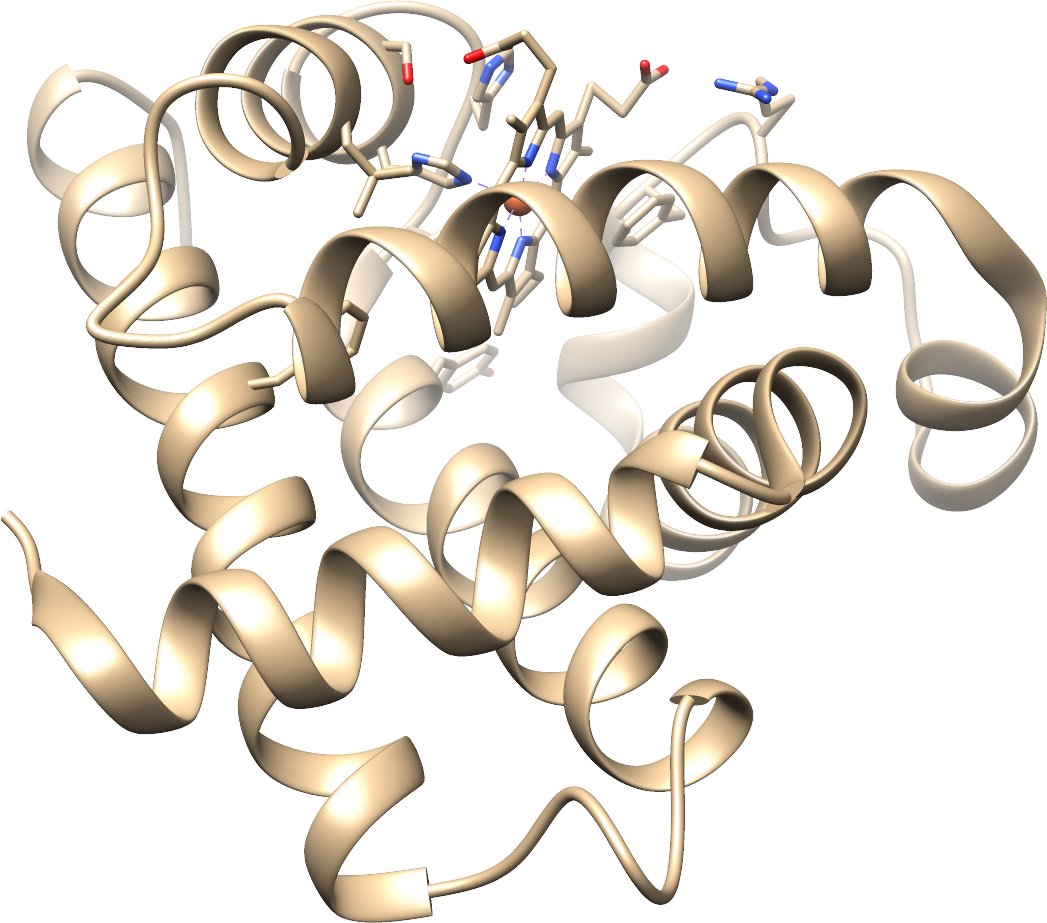}
(b)\hspace{-1.5em}%
\includegraphics[width=0.5\linewidth]{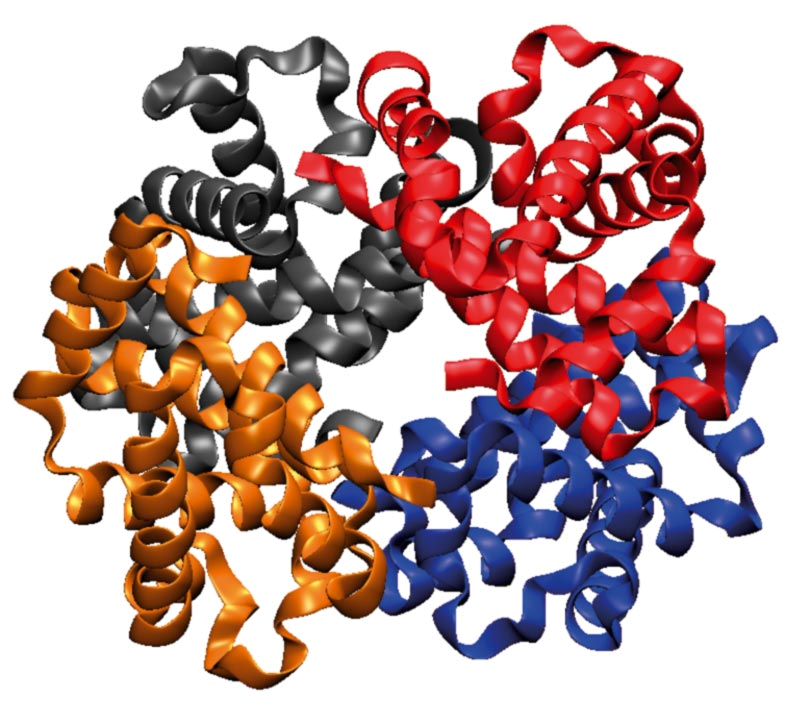}

Myoglobin is the oxygen carrier in muscles \cite{Kendrew1960} \pdbref{1mbn} (a). 
Hemoglobin is the oxygen carrying complex found in all red blood cells, it is homologous to myoglobin but consists of four subunits (b). It has an interesting allosteric mechanism that enhances its affinity for oxygen, and the same motion occurs in the myoglobin monomer. 

Snapshots of time-resolved X-ray crystallography in myoglobin 
can be seen in Fig.\@ 7 of \citet{Schotte2004}. 
These figures show the basis of the allosteric mechanism, where carbon monoxide (CO) was bound in stead of the oxygen (O$_2$). We first see a response from the photo-induced breakage of the  CO iron bond, and subsequent moving away of the CO. At about 30 nanoseconds we can see how the unbound CO moves further away. At 3 microseconds we see the CO slowly returning and re-binding to the iron. More detail, and explanation of the colouring scheme used in \cite{Schotte2004}. 

In hemoglobin, the motion of the histidine (Hi93 in myoglobin), shifts the orientation of one alpha-helix, which sits at the interface with the other hemoglobin monomers. This causes a similar shift in the other monomers, pulling on the same histidine, and distorting the shape of the heme group so oxygen binding becomes more favourable. 

\end{bgreading}

For NMR there is a one-to-one relation between signals measured and particular atoms, bonds or angles. The width of these peaks can vary, and this is usually entirely due to (local) structure and dynamics of the molecule, i.e.\@ how the atoms are oriented and how much they move.
In case of particularly dynamic molecules, one should realize that some atoms may alternatively be close to two different parts of the molecule. For example, a dynamic protein loop that has two conformations. Both distances will be short enough to yield a measurable NOE intensity; but no structure exists that can satisfy both short distances at the same time. For this, and other reasons, NMR experimental data is typically used to generate an ensemble of solution structures instead of a single one as is done for X-ray. One may think of this as reflecting the innate dynamics of the protein. However, be aware that sparsity of data may result in an under-defined structure, which will also yield larger variations in the ensemble generated, but this does not arise from dynamics.

X-ray crystallography and cryo-EM provide very detailed, but intrinsically static pictures of protein structures. The dynamics of protein structures are poorly represented by this static view. NMR and other spectroscopic techniques help remedy this, and these are often used in combination with molecular simulations, which we will cover in \chref{ChIntroDyn} and subsequent chapters on thermodynamics and simulations.

\section{Key points}
\begin{compactitem}
\item 3D coordinates of the protein (PDB) are \textbf{not the primary experimental data}. 
\item For X-ray crystallography: 
\begin{compactitem}
\item Electron density maps are also not the primary data
\item \textbf{Diffraction patterns} are the primary data
\end{compactitem}
\item For NMR:
\begin{compactitem}
\item Distances and angles are also not the primary data
\item \textbf{Spectra and intensities} are the primary data
\end{compactitem}
\item Everything else is (at least partly) \textbf{based on modelling}
\item For the other techniques mentioned, this dependence on modelling is even stronger
\item Proteins are not static, they are \textbf{dynamic}
\end{compactitem}

\section{Recommended further reading}
\label{sec:ChDetVal:reading}
\begin{compactitem}


\item \citet{BrandenTooze} -- ``Introduction to protein structure'' for a broader general introduction to protein structure and structure determination.
\item \citet{Atkins2002} -- ``Physical Chemistry'' for a more in-depth on structure determination of biological macromolecules, and other experimental approaches to elucidate functional, structural and chemical properties.
\item \citet{Giacovazzo2011} -- ``Fundamentals of Crystallography'' for an advanced account of modern crystallography, including the mathematical details of different approaches and techniques, highly recommended by practising crystallographers.
\item \citet{Shen2018} \url{https://cryoem101.org/chapter-1/} -- an accessible introduction to the experimental and data processing basics of cryo-EM.
\item \citet{Teilum2017} ``(S)Pinning down protein interactions by NMR''
\item \citet{Kwan2011} ``Macromolecular NMR for the non-spectroscopist''

\end{compactitem}


\section*{Author contributions}
{\renewcommand{\arraystretch}{1}
\begin{tabular}{@{}ll}
\ACtxt: &   HM, BS, HI, KAF \\
\ACfig: &   HM, HI, JG, KW, KAF \\
\ACref: &   HM, HI, KW, KAF \\
\ACproof:&  BS, SA \\
\ACfb:  &   JG, KW \\
\ACeds: &   HM, SA, KAF
\end{tabular}}

\noindent
The authors thank \AR~\ARid{} for creating \figref{ChStrucDet-crystal}.

\mychapbib

\clearpage

\cleardoublepage

\end{document}